\documentclass[prc,twocolumn,showkeywords,showpreprintnumbers,floatfix]{revtex4}
\usepackage{amsmath,amssymb,graphicx,bm}
\usepackage{color}
\usepackage{ulem}
\usepackage{graphicx}
\usepackage{dcolumn}

\newcommand{\vek}[1]{\bm{\mathrm{#1}}}
\DeclareMathOperator{\artanh}{artanh}
\DeclareMathOperator{\tr}{Tr}
\DeclareMathOperator{\im}{Im}
\DeclareMathOperator{\re}{Re}
\newcommand{\qmax}{q_{\text{max}}}
\newcommand{\vlowk}{V_{\text{low}\,k}}

\newcommand{\fmi}{\, \text{fm}^{-1}}

\newcommand{\fmithree}{\, \text{fm}^{-3}}

\newcommand{\Gtemp}{\mathcal{G}}
\newcommand{\Kernel}{\mathcal{K}}
\newcommand{\Tmatrix}{\mathcal{T}}
\newcommand{\calT}{\mathcal{T}}
\newcommand{\Vind}{V_{\text{ind}}}
\newcommand{\tot}{\text{tot}}
\newcommand{\corr}{\text{corr}}
\newcommand{\half}{\frac{1}{2}}
\newcommand{\CG}[6]{C_{#1 #2 #3 #4}^{#5 #6}}
\newcommand{\Sec}[1]{Sec.~\ref{#1}}

\newcommand{\Ref}[1]{Ref.~\cite{#1}}

\newcommand{\Eq}[1]{Eq.~(\ref{#1})}
\newcommand{\Eqs}[1]{Eqs.~(\ref{#1})}
\newcommand{\Fig}[1]{Fig.~\ref{#1}}
\newcommand{\Figs}[1]{Figs.~\ref{#1}}

\begin{document}

\title{Screening and anti-screening of the pairing interaction in low-density
  neutron matter}
\author{S. Ramanan} \email{suna@physics.iitm.ac.in}
\affiliation{Department of Physics, Indian Institute of Technology Madras,
  Chennai - 600036, India} \author{M. Urban}
\email{urban@ipno.in2p3.fr} \affiliation{Institut de Physique
  Nucl\'eaire, CNRS-IN2P3, Univ. Paris-Sud, Universit\'e Paris-Saclay, 91406 Orsay
  cedex, France} 
\begin{abstract}
We study pairing in low-density neutron matter including the screening
interaction due to the exchange of particle-hole and RPA
excitations. As bare force we employ the effective low-momentum
interaction $\vlowk$, while the Fermi-liquid parameters are taken from
a phenomenological energy density functional (SLy4) which correctly
reproduces the equation of state of neutron matter. At low density, we
find screening, i.e., pairing is reduced, while at higher densities,
we find anti-screening, i.e., pairing is enhanced. This
  enhancement is mostly due to the strongly attractive Landau
  parameter $f_0$. We discuss in detail the critical temperature $T_c$
  in the limit of low densities and show that the suppression of $T_c$
  predicted by Gor'kov and Melik-Barkhudarov can only be reproduced if
  the cutoff of the $\vlowk$ interaction is scaled with the Fermi
  momentum. We also discuss the effect of non-condensed pairs on the
  density dependence of $T_c$ in the framework of the
  Nozi\`eres-Schmitt-Rink theory.
\end{abstract}

\keywords{Neutron matter, pairing, screening}

\maketitle

%%%%%%%%%%%%%%%%%%%%%%%%%%%%%%%%%%%%%%%%%%%%%%%%%%%%%%%%%%%%%%%%%%%%%%%%%%%%%%%%
\section{Introduction}

Neutron stars provide a unique laboratory with an interplay of a wide
range of phenomena. The physics of the inner crust of neutron stars,
where a dilute gas of unbound neutrons coexists with nuclear clusters,
is particularly interesting~\cite{liv-rev}. In this work, we focus on
the neutron gas, since its superfluid properties are crucial for the
understanding of astrophysical observables such as pulsar glitches or
neutron-star cooling. Glitches are the observed sudden increase in the
rotational frequency of the pulsars, followed by a long relaxation
time and usually they are linked to the
  neutron superfluidity in the inner 
  crust~\cite{anderson-itoh,pines-alpar,haskell-sedrakian}, in
particular, to the unpinning of the vortices. After the initial rapid
cooling via neutrino emissions, the cooling rate of the neutron star
is very dependent on the physics of the crust. The superfluidity of
the neutrons in the crust of the star strongly suppresses the specific
heat and hence influences the cooling rate
\cite{yakovlev-pethick,Fortin2010}. In addition, neutron superfluidity 
allows for novel neutrino emission processes via Cooper pair breaking 
and formation that affect the cooling rate of the star close to the transition 
temperature~\cite{page-lattimer-prakash-steiner}.

Even the modelling of uniform matter is theoretically very challenging
due to the uncertainties in the nuclear interactions.  In neutron
stars, the attractive interaction is provided by the two-body
interaction, and the most important channels for neutron pairing turn
out to be the ${}^1S_0$ channel at low densities and therefore
occuring in the inner crust, while in the core, the neutrons pair in
the triplet ${}^3P_2-{}^3F_2$ channel. Protons can also pair, although
a description of proton superfluidity is complicated by the asymmetry
of matter and the resulting coupling of the protons to the denser
background~\cite{gezerlis-pethick-schwenk}. In addition to being
crucial for the physics of neutron stars, pairing between nucleons
plays a very important role in the spectra of finite nuclei, as well
as in description of neutron rich nuclei close to the drip line.

A reliable description of pairing at all densities in infinite matter
is still an open question, although the superfluidity in stars has
been studied since the early work of Migdal~\cite{migdal} and Ginzburg
and Kirzhnits~\cite{ginzburg64,ginzburg69} and is needed to
  explain observations such as the long relaxation time after a glitch
  \cite{baym-pethick-pines-ruderman}. For a recent review, the reader
is referred to~\cite{sedrakian2018}. The simplest starting point for
the study of pairing is the superfluid gap equation within the BCS
approximation that uses the free-space two-nucleon interaction as
input and a free spectrum for the single-particle energies. However,
there is enough evidence that one needs to go beyond this
approximation~\cite{gezerlis-pethick-schwenk,dean-jensen,Srinivas2016,Drischler2016,Papakonstantinou2017,Rios2017,Rios2018}. Medium
corrections to the single-particle energy and to the free-space
interaction change the gap drastically.

In this work, we re-visit the issue of building
an induced interaction that will modify the free space two-body
interaction responsible for pairing in the $^1S_0$ channel in uniform
neutron matter. In the past, several attempts have been made to
include medium corrections to the
interaction~\cite{Wambach1993,Schulze1996,Shen2003,Shen2005,Cao2006}.
Most of these calculations use many-body methods analogous
  to the well-known example of screening in an electron gas
  \cite{FetterWalecka}, subject to various approximations. Because of the exponential
dependence of the gap on the interaction, the final results are always
affected by the details. In view of the persistent
  uncertainties, of some mistakes in \Ref{Shen2005} (see
  \Ref{Cao2006}), and of the simplifying approximation made in
  \Ref{Cao2006} to replace the \mbox{3 particle}$-$\mbox{1 hole}
  (3p1h) matrix element entering the induced interaction by its
  average value, we believe that this problem has not yet been fully
  solved, even within the given theoretical framework.
  
As input, we use the free-space renormalized two-body interaction,
$\vlowk$, evolved from the AV$_{18}$ two-body potential. The same
interaction is also used for the 3p1h couplings entering the induced
interaction. The main advantage of using $\vlowk$ is that
non-perturbative features present in the bare interaction, such as the
short-range repulsion that arises from the hard-core and the repulsive
tensor, are softened. The low-momentum effective interaction depends
on the renormalization scale (or cutoff) $\Lambda$, while the
free-space two-body observables such as scattering phase shifts and
energies are independent of $\Lambda$. However, in principle the
renormalization group running generates also induced three- and
higher-body forces. In addition, in a many-body calculation, one
usually resorts to approximations, which may not hold for all
situations. Therefore, when only the free-space evolved two-body
interaction is used as input in a many-body calculation, the results
may depend on the cutoff and this dependence gives not only an
estimate of the importance of the missing $3N$ force but also
indicates the importance of the missing many-body terms that may
become relevant~\cite{hebeler-schwenk2010}.

For the induced interaction, except at extremely low densities, it is
necessary to go beyond the exchange of simple particle-hole
excitations. Following \Ref{Cao2006}, we sum the particle-hole bubble
series (random-phase approximation, RPA) within the Landau
approximation and keep only the lowest order ($L = 0$) Landau
parameters in the particle-hole interaction. In \Ref{Cao2006},
  as in preceding studies \cite{Schulze1996,Shen2005}, the Landau
  parameters were computed microscopically, including the induced
  interaction in a self-consistent manner (so-called Babu-Brown theory
  \cite{Babu1973}). However, the resulting Landau parameters, in
  particular $F_0$, were much smaller than what one obtains from
  phenomenological energy-density functionals such as the Skyrme SLy4
  or the Gogny D1N parameterizations, which have both been fitted to the
  neutron-matter equation of state. Therefore, we follow a more
  pragmatic but probably more reliable strategy here, namely to
  determine the Fermi-liquid parameters (Landau parameters and
  effective mass) directly from these phenomenological interactions.

The medium corrected interaction is then used in the BCS gap equation
and the transition temperature is calculated. We note that our results
show screening at low densities and anti-screening at high
densities. This is different from the results of Cao et al. in
\Ref{Cao2006}, where they predict screening for all densities. Our
results for screening, e.g., \Fig{fig:Tcratio}, are compatible with
Quantum Monte-Carlo (QMC)
results~\cite{Abe2009,Gandolfi2008,Gezerlis2010} which rule out the
extremely strong screening predicted in earlier calculations
\cite{Wambach1993}. Unfortunately, QMC results are not available in
the density range where we find anti-screening.

Apart from the induced interaction, there are other effects that
  may modify the BCS results for the transition temperature. If the
  Fermi momentum $k_F$ lies approximately between
  $1/|a|\sim 0.05$ fm$^{-1}$ and $1/r_e\sim 0.4$
  fm$^{-1}$, where $a$ is the neutron-neutron ($nn$) scattering
  length and $r_e$ the effective range, neutron matter is in a
  strong-coupling situation, in which pair correlations appear already
  in the normal phase and modify the critical temperature $T_c$
  \cite{Nozieres1985}. This effect is crucial for the understanding of
  the BCS-BEC crossover as it exists in ultracold Fermi gases or in
  symmetric nuclear matter \cite{Jin2010}, where one can pass from
  Cooper pairs to a Bose-Einstein condensate (BEC) of dimers
  (deuterons). For a recent review article, see
  \cite{Strinati2018}. The large value of $|a|$
  indicates that the $nn$ interaction is almost able to produce a
  bound state, and in low-density neutron matter the $nn$ Cooper-pair
  wave function indeed looks almost like a bound-state wave function
  \cite{Matsuo2006,Margueron2007,Sun2010}. In fact, one can reach a
  situation similar to the unitary limit, which is the case of a
  contact interaction with $|a|\to\infty$ (i.e.,
  $1/|a| \ll k_F\ll 1/r_e$). The relevance of BEC-BCS
  cross-over physics for the description of dilute neutron matter was
  pointed out in many works, e.g.
  \cite{Abe2009,Gezerlis2010,Matsuo2006,Margueron2007,Sun2010,Ramanan2013}.

Note that, although the strong-coupling situation is only reached at
densities below $\sim 0.01$ times nuclear saturation density, it is
phenomenologically relevant. Neutron matter with such low densities is
present between the clusters in the inner crust of neutron stars at
average baryon densities just above the neutron-drip density of $\sim
2.5\times 10^{-4}$ fm$^{-3}$ \cite{Negele1973,Baldo2007}. Since in this
region the dilute neutron gas fills almost the entire volume, it
represents a sizable contribution to the average baryon density even
if its density is a few thousand times smaller than the density inside
the clusters.

In the unitary limit, the Nozi\`eres-Schmitt-Rink (NSR) theory of pair
correlations in the normal phase \cite{Nozieres1985} predicts a
reduction of the transition temperature $T_c$ from the BCS result
$\sim 0.5\,E_F$ ($E_F$ being the Fermi energy) to $\sim 0.22\,E_F$
\cite{Sa-de-Melo1993}. These numbers do not include screening effects,
but as shown recently \cite{Pisani2018}, the inclusion of screening on
top of the NSR effect leads to good agreement with results from
experiments with ultracold atoms. In a previous work
\cite{Ramanan2013}, we had studied neutron matter in the framework of
the NSR theory. In the present paper, we will extend that work to see
how the NSR correction is changed by the induced interaction.

This paper is organised as follows: in \Sec{sec:formalism}, we
re-visit the BCS gap equation and set up the induced interaction. The
effect of the induced interaction on the transition temperature is
discussed in \Sec{sec:RPA}. At low densities, one expects a reduction
in $T_c$ by a factor of $(4e)^{-1/3}$, which is the
Gor'kov-Melik-Barkhudarov (GMB) result \cite{Gorkov1961}, and this
region is studied in detail in \Sec{sec:low-density-limit}. Finally,
we turn our attention to the correlations within the NSR approach in
\Sec{sec:NSR}. A summary of our results is presented in
\Sec{sec:conclude}. Some of the details of the calculations have been
moved to appendices to facilitate ease of reading. Numerical
  results for the matrix elements of the screened pairing interaction
  are provided in the supplemental material \cite{suppl}.

%%%%%%%%%%%%%%%%%%%%%%%%%%%%%%%%%%%%%%%%%%%%%%%%%%%%%%%%%%%%%%%%%%%%%%%%%%%%%%%%
\section{Formalism}
\label{sec:formalism}
%%%%%%%%%%%%%%%%%%%%%%%%%%%%%%%%%%%%%%%%%%%%%%%%%%%%%%%%%%%%%%%%%%%%%%%%%%%%%%%%
\subsection{Gap equation and induced interaction}
%%%%%%%%%%%%%%%%%%%%%%%%%%%%%%%%%%%%%%%%%%%%%%%%%%%%%%%%%%%%%%%%%%%%%%%%%%%%%%%%
In BCS theory, the $^1S_0$ pairing gap $\Delta$ in neutron matter is
determined by the gap equation
\begin{equation}
\Delta(k) = -\frac{2}{\pi}\int_0^\infty \!dq\, q^2
V_0(k,q) \frac{\Delta(q)\tanh\big(\frac{E(q)}{2T}\big)}{2E(q)}\,.
\label{gapeq}
\end{equation}
Here, $V_0(k,q) = \langle k|V_{^1S_0}|q\rangle$ denotes the
$nn$ interaction in the $^1S_0$ partial wave for in- and
outgoing momenta $q$ and $k$, $E_q =
\sqrt{(\epsilon(q)-\mu)^2+\Delta(q)^2}$ is the quasiparticle energy
with $\epsilon(q) = q^2/2m^*$, $m^*$ is the neutron effective mass,
$\mu$ is the effective chemical potential including the mean-field
energy shift, and $T$ is the temperature. Except in some range of low
densities, neutron matter is in the weak-coupling limit, in the sense
that $\Delta(k_F) \ll \mu$, implying $\mu \approx k_F^2/2m^*$, with
the Fermi momentum $k_F = (3\pi^2\rho)^{1/3}$ determined by the
neutron number density $\rho$. Equation (\ref{gapeq}) with
  $\vlowk$ as $nn$ interaction has been solved, e.g., in
  \Ref{Kuckei2003}.

The critical temperature $T_c$ is the highest temperature for which
\Eq{gapeq} has a non-trivial solution.  At $T=T_c$, one can
neglect $\Delta(q)$ in $E(q)$, so that \Eq{gapeq} reduces to a linear
eigenvalue equation
\begin{equation}
\phi(k) = -\frac{2}{\pi}\int_0^\infty \!dq\, q^2 V_0(k,q)
\frac{\tanh\big(\frac{\xi(q)}{2T_c}\big)}{2\xi(q)}\phi(q)\,,
\label{eq:Tc}
\end{equation}
with $\xi(q) = \epsilon(q)-\mu$. We will also write this as
$|\phi\rangle = \Kernel|\phi\rangle$. Hence, in order to find $T_c$, we
diagonalize the integral operator with the kernel
\begin{equation}
  \Kernel(k,q) = 
  -V_0(k,q)\frac{\tanh\big(\frac{\xi(q)}{2T}\big)}{2\xi(q)}\,,
  \label{eq:kernel}
\end{equation}
and $T_c$ is the temperature where the largest eigenvalue is equal to
unity. In weak coupling, $T_c$ is directly related to the gap at $T=0$
by $T_c = 0.57\, \Delta_{T=0}(k_F)$.

\begin{figure}
\includegraphics[scale=0.9]{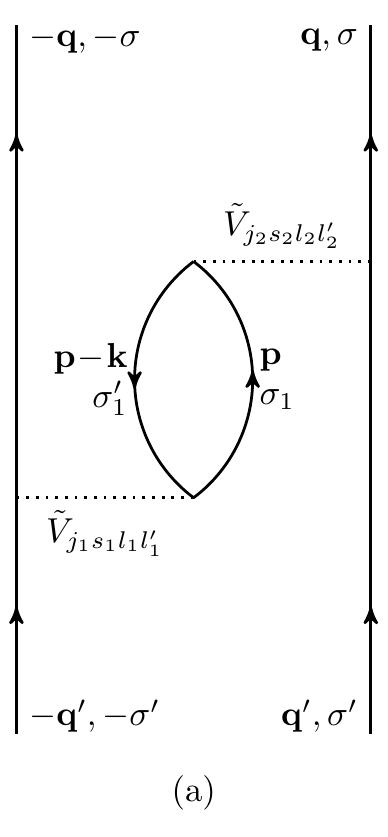}\hspace{1cm}
\includegraphics[scale=0.9]{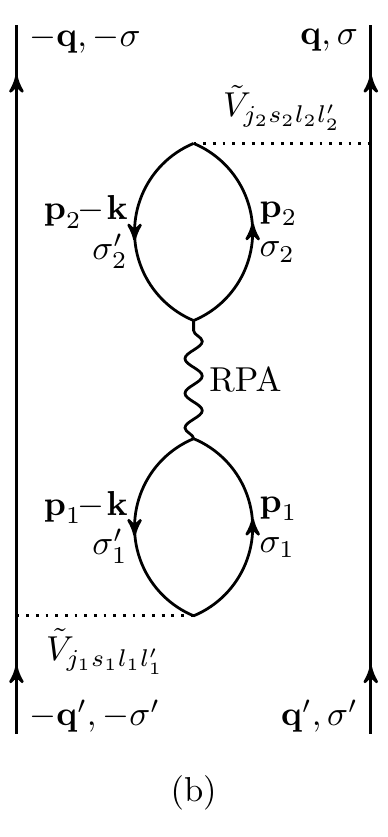}
\caption{\label{fig:diagrams}Feynman diagrams representing the induced
  interaction. The wiggly line in diagram (b) is meant to include the
  RPA bubble summation.}
\end{figure}

It is widely accepted that an important correction to BCS theory
consists in adding to the bare interaction in \Eq{gapeq} the
contribution of the induced interaction $\Vind$ due to the exchange of
density and spin-density fluctuations. In particular, in the weakly
interacting limit, this leads to the famous Gor'kov-Melik-Barkhudarov
(GMB) correction, which reduces the gap and the critical temperature
by approximately a factor of two compared to the BCS result
\cite{Gorkov1961}. In terms of Feynman diagrams, this correction can
be represented as in \Fig{fig:diagrams} (a).
%%%%%%%%%%%%%%%%%%%%%%%%%%%%%%%%%%%%%%%%%%%%%%%%%%%%%%%%%%%%%%%%%%%%%%%%%%%%%%%%
%%%%%%%%%%%%%%%%%%%%%%%%%%%%%%%%%%%%%%%%%%%%%%%%%%%%%%%%%%%%%%%%%%%%%%%%%%%%%%%%
Note that the dotted interaction lines are meant to represent the
antisymmetrized interaction. This is very important since the dominant
$^1S_0$ interaction acts only between neutrons of opposite spin and
therefore cannot contribute to the shown diagram. However, if the
outgoing lines are exchanged in both the interaction vertices, one
obtains a diagram to which it contributes.

In nuclear physics, except at extremely low density (see
\Sec{sec:low-density-limit}), one is never in a weakly interacting
regime. Therefore, the simple particle-hole bubble exchanged in
\Fig{fig:diagrams} (a) is modified by the residual particle-hole
interaction as shown in \Fig{fig:diagrams} (b). The wiggly line
representing the particle-hole interaction is meant to include the RPA
bubble summation to all orders.

Throughout this article, ``diagram (a)'' and ``diagram (b)'' refer to
the diagrams shown in \Fig{fig:diagrams} (a) and (b). When calculating
the diagrams, we make the usual approximation to neglect the energy
transfer (static approximation) which can be justified by the
observation that the most important contribution to pairing comes from
scattering of particles near the Fermi surface, so that all in- and
outgoing particles have energies close to the Fermi energy $\epsilon_F
= k_F^2/2m^*$.

%%%%%%%%%%%%%%%%%%%%%%%%%%%%%%%%%%%%%%%%%%%%%%%%%%%%%%%%%%%%%%%%%%%%%%%%%%%%%%%%
\subsection{Diagram (a): single-bubble exchange}
\begin{figure}
\includegraphics[scale=0.9]{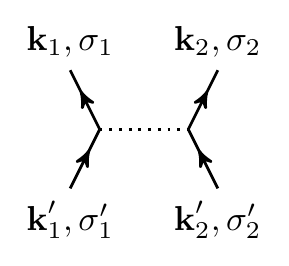}\hspace{1cm}
\includegraphics[scale=0.9]{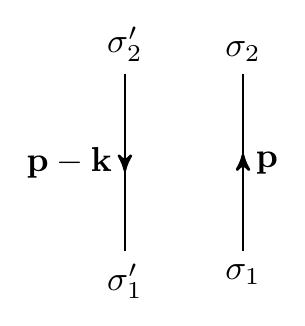}
\caption{\label{fig:Feynman-elements} Elements of Feynman diagrams to
  clarify the notation. Left: particle-particle interaction. Right:
  particle-hole propagator.}
\end{figure}

%%%%%%%%%%%%%%%%%%%%%%%%%%%%%%%%%%%%%%%%%%%%%%%%%%%%%%%%%%%%%%%%%%%%%%%%%%%%%%%%
Let us first discuss the vertices coupling the particles to the
particle-hole excitation, represented as dotted lines in
\Fig{fig:diagrams}. We assume a general (possibly non-local)
interaction which is expanded in partial waves. Using the notation of
the left part of \Fig{fig:Feynman-elements},
%%%%%%%%%%%%%%%%%%%%%%%%%%%%%%%%%%%%%%%%%%%%%%%%%%%%%%%%%%%%%%%%%%%%%%%%%%%%%%%%
%%%%%%%%%%%%%%%%%%%%%%%%%%%%%%%%%%%%%%%%%%%%%%%%%%%%%%%%%%%%%%%%%%%%%%%%%%%%%%
the partial-wave expansion of the interaction reads
\begin{multline}
\langle \vek{k}_1,\sigma_1; \vek{k}_2,\sigma_2|V
  |\vek{k}'_1,\sigma'_1; \vek{k}'_2,\sigma'_2\rangle =\\
    \sum_{s,m_s,m'_s} \sum_{l,l',m_l} \sum_{j}
      \CG{\half}{\sigma_1}{\half}{\sigma_2}{s}{m_s}
      \CG{\half}{\sigma'_1}{\half}{\sigma'_2}{s}{m'_s}
      \CG{l}{m_l}{s}{m_s}{j}{m_j}
      \CG{l'}{m'_l}{s}{m'_s}{j}{m_j}\\ \times 
      (4\pi)^2 i^{l'-l} Y^*_{lm_l}(\Omega_{\vek{Q}})Y_{l'm'_l}(\Omega_{\vek{Q}'})
      \langle Q|V_{sll'j}|Q'\rangle\,,
      \label{eq:partialwaves}
\end{multline}
with
\begin{gather}
\vek{Q} = \frac{\vek{k}_1-\vek{k}_2}{2}\,,\quad
\vek{Q}' = \frac{\vek{k}'_1-\vek{k}'_2}{2}\,,\\
  m'_l = m_l+m_s-m'_s\,,\quad m_j = m_l+m_s\,.
\end{gather}
For the Clebsch-Gordan coefficients, we follow the notation of the
book by Varshalovich \cite{Varshalovich}.

Then it is straight-forward to obtain for the diagram (a) the
following expression [the factor $(-1)$ comes from the closed Fermion
  loop]:
\begin{widetext}
\begin{multline}
V_{a}(q,q') = (-1)\frac{1}{4\pi}
  \sum_{\sigma\sigma'}
    \CG{\half}{\sigma}{\half}{-\sigma}{0}{0}
    \CG{\half}{\sigma'}{\half}{-\sigma'}{0}{0}
  \int \frac{d\Omega_{\vek{q}}}{4\pi}
  \int \frac{d\Omega_{\vek{q}'}}{4\pi}
  \int \frac{d^3p}{(2\pi)^3}\,
    \frac{n(\vek{p}-\vek{k})-n(\vek{p})}
      {\epsilon(\vek{p})-\epsilon(\vek{p}-\vek{k})}\\ \times
  \sum_{s_1 m_{s1}}\sum_{s_2 m_{s2}}
  \sum_{l_1l'_1m_{l1}}\sum_{l_2,l'_2,m_{l2}}
  \sum_{j_1j_2}
    \CG{\half}{\sigma_1}{\half}{-\sigma}{s_1}{m_{s1}}
    \CG{\half}{\sigma'_1}{\half}{-\sigma'}{s_1}{m'_{s1}}
    \CG{\half}{\sigma}{\half}{\sigma'_1}{s_2}{m_{s2}}
    \CG{\half}{\sigma'}{\half}{\sigma_1}{s_2}{m'_{s2}}
    \CG{l_1}{m_{l1}}{s_1}{m_{s1}}{j_1}{m_{j1}}
    \CG{l'_1}{m'_{l1}}{s_1}{m'_{s1}}{j_1}{m_{j1}}
    \CG{l_2}{m_{l2}}{s_2}{m_{s2}}{j_2}{m_{j2}}
    \CG{l'_2}{m'_{l2}}{s_2}{m'_{s2}}{j_2}{m_{j2}}\\ \times
    (4\pi)^4 i^{l'_1-l_1+l'_2-l_2} 
      Y^*_{l_1m_{l1}}(\Omega_{\vek{Q}_1}) 
      Y_{l'_1m'_{l1}}(\Omega_{\vek{Q}'_1})
      Y^*_{l_2m_{l2}}(\Omega_{\vek{Q}_2})
      Y_{l'_2m'_{l2}}(\Omega_{\vek{Q}'_2})
    \langle Q_1|\tilde{V}_{s_1l_1l'_1j_1}|Q'_1\rangle
    \langle Q_2|\tilde{V}_{s_2l_2l'_2j_2}|Q'_2\rangle\,,
    \label{eq:diaga}
\end{multline}
\end{widetext}
with the following abbreviations:
\begin{gather}
  \vek{k} = \vek{q}-\vek{q}'\,,\label{definition-k-a}\displaybreak[0]\\
\begin{split}
  &\vek{Q}_1 = \frac{\vek{q}+\vek{p}}{2}\,,\quad
  \vek{Q}'_1 = \frac{\vek{q}'-\vek{k}+\vek{p}}{2}\,,\\
  &\vek{Q}_2 = \frac{\vek{q}+\vek{k}-\vek{p}}{2}\,,\quad
  \vek{Q}'_2 = \frac{\vek{q}'-\vek{p}}{2}\,,
  \label{definition-Q-a}
\end{split}\displaybreak[0]\\
  \sigma_1 = m_{s1}+\sigma\,,\quad
  \sigma'_1 = m_{s2}-\sigma\,,\label{definition-sigma-a}\displaybreak[0]\\
  m'_{s1} = m_{s2}-\sigma-\sigma'\,,\quad
  m'_{s2} = m_{s1}+\sigma+\sigma'\,,\label{definition-ms-a}\displaybreak[0]\\
  m'_{l1} = m_{l1}+m_{s1}-m'_{s1}\,,\quad
  m'_{l2} = m_{l2}+m_{s2}-m'_{s2}\,,\label{definition-ml-a}\displaybreak[0]\\
  m_{j1} = m_{l1}+m_{s1}\,,\quad
  m_{j2} = m_{l2}+m_{s2}\,.\label{definition-mj-a}
\end{gather}

The tilde in $\tilde{V}$ indicates that the matrix element is
antisymmetrized, i.e., multiplied by a factor of two in the surviving
channels. For the occupation numbers $n(\vek{p})$ and $n(\vek{k}-\vek{p})$
entering the integral in \Eq{eq:diaga}, we use the step function
$n(\vek{p}) = \theta(k_F-p)$, which is a very good approximation as
long as we are in the weak-coupling limit ($\Delta,T\ll \mu$). Notice
that then
\begin{equation}
  \lim_{k\to 0}\frac{n(\vek{p}-\vek{k})-n(\vek{p})}
      {\epsilon(\vek{p})-\epsilon(\vek{p}-\vek{k})} =
  m^*\delta(p-k_F)\,,
\end{equation}
which is useful when evaluating \Eq{eq:diaga} for $q = q'$,
especially in the case $q=q'=0$.

%%%%%%%%%%%%%%%%%%%%%%%%%%%%%%%%%%%%%%%%%%%%%%%%%%%%%%%%%%%%%%%%%%%%%%%%%%%%%%%%
\subsection{Separation of $S=0$ and $S=1$ contributions}
%%%%%%%%%%%%%%%%%%%%%%%%%%%%%%%%%%%%%%%%%%%%%%%%%%%%%%%%%%%%%%%%%%%%%%%%%%%%%%%%
It is instructive to split \Eq{eq:diaga} into contributions from
particle-hole excitations having total spin $S=0$ (density waves) and
$S=1$ (spin-density waves). In order to do this, consider the
particle-hole propagator shown in the right part of
\Fig{fig:Feynman-elements}, which is given by
$G_0(p)G_0(p-k)\delta_{\sigma_1\sigma_2}\delta_{\sigma'_1\sigma'_2}$
with $G_0$ the uncorrelated single-particle Green's
function. This expression appears also in diagram (a) if we formally
introduce a summation over $\sigma_2$ and $\sigma'_2$. The spin part
can be decomposed using the completeness relation of the Pauli
matrices $\vek{\sigma}$
\begin{equation}
\delta_{\sigma_1\sigma_2}\delta_{\sigma'_1\sigma'_2} =
  \tfrac{1}{2}(\delta_{\sigma_1\sigma'_1}\delta_{\sigma_2\sigma'_2}
    +\vek{\sigma}_{\sigma'_1\sigma_1}\cdot\vek{\sigma}_{\sigma_2\sigma'_2})\,,
\label{eq:completeness-pauli}
\end{equation}
where the two terms correspond, respectively, to $S=0$ and
$S=1$. Likewise, this decomposition can also be written in terms of
Clebsch-Gordan coefficients as
\begin{multline}
\delta_{\sigma_1\sigma_2}\delta_{\sigma'_1\sigma'_2} = \\
  \sum_{S,m_S} 
    (-1)^{\half-\sigma'_1} \CG{\half}{\sigma_1}{\half}{-\sigma'_1}{S}{m_S}
    (-1)^{\half-\sigma'_2} \CG{\half}{\sigma_2}{\half}{-\sigma'_2}{S}{m_S}\,.
\label{eq:completeness-cg}
\end{multline}
In the calculation of $V_{a}(q,q')$, it is clear that in the $S=1$
case each of the three spin projections $m_S$ of the particle-hole
excitation must give the same contribution. We can therefore compute
the $S=1$ contribution by restricting ourselves to the $m_S=0$ term,
or, equivalently, by keeping only the Pauli matrix $\sigma_z$ in the
second term of \Eq{eq:completeness-pauli}, and multiplying the result
by three. This amounts to the replacement
\begin{equation}
\delta_{\sigma_1\sigma_2}\delta_{\sigma'_1\sigma'_2}\to 
  \tfrac{1}{2}\delta_{\sigma_1\sigma'_1}\delta_{\sigma_2\sigma'_2}
  \left[1+3\,(-1)^{1-\sigma_1-\sigma_2}\right]\,.
\end{equation}

In this way, we arrive at an alternative expression for diagram (a):
\begin{widetext}
\begin{multline}
V_{a}(q,q') = (-1)\frac{1}{4\pi}
    \sum_{\sigma\sigma'}
    \CG{\half}{\sigma}{\half}{-\sigma}{0}{0}
    \CG{\half}{\sigma'}{\half}{-\sigma'}{0}{0}
    \int \frac{d\Omega_{\vek{q}}}{4\pi}
    \int \frac{d\Omega_{\vek{q}'}}{4\pi}
    \int \frac{d^3p}{(2\pi)^3}\,
    \frac{n(\vek{p}-\vek{k})-n(\vek{p})}
      {\epsilon(\vek{p})-\epsilon(\vek{p}-\vek{k})}\\ \times
    \sum_{s_1 m_{s1}}\sum_{s_2 m_{s2}}
    \sum_{l_1l'_1m_{l1}}\sum_{l_2,l'_2,m_{l2}}\sum_{j_1j_2}
    \CG{\half}{\sigma_1}{\half}{-\sigma}{s_1}{m_{s1}}
    \CG{\half}{\sigma_1}{\half}{-\sigma'}{s_1}{m'_{s1}}
    \CG{\half}{\sigma}{\half}{\sigma_2}{s_2}{m_{s2}}
    \CG{\half}{\sigma'}{\half}{\sigma_2}{s_2}{m'_{s2}}
    \CG{l_1}{m_{l1}}{s_1}{m_{s1}}{j_1}{m_{j1}}
    \CG{l'_1}{m'_{l1}}{s_1}{m'_{s1}}{j_1}{m_{j1}}
    \CG{l_2}{m_{l2}}{s_2}{m_{s2}}{j_2}{m_{j2}}
    \CG{l'_2}{m'_{l2}}{s_2}{m'_{s2}}{j_2}{m_{j2}}\\ \times
    (4\pi)^4 i^{l'_1-l_1+l'_2-l_2} 
      Y^*_{l_1m_{l1}}(\Omega_{\vek{Q}_1}) 
      Y_{l'_1m'_{l1}}(\Omega_{\vek{Q}'_1})
      Y^*_{l_2m_{l2}}(\Omega_{\vek{Q}_2})
      Y_{l'_2m'_{l2}}(\Omega_{\vek{Q}'_2})
    \langle Q_1|\tilde{V}_{s_1l_1l'_1j_1}|Q'_1\rangle
    \langle Q_2|\tilde{V}_{s_2l_2l'_2j_2}|Q'_2\rangle\\ \times
    \frac{1}{2}\left[1+3\,(-1)^{1-m_{s1}-m_{s2}}\right]\,,
    \label{eq:diagaa}
\end{multline}
\end{widetext}
with the same abbreviations $\vek{k}$, $\vek{Q}_i$, $\vek{Q}'_i$,
$m'_{li}$, and $m_{ji}$ as before [Eqs. (\ref{definition-k-a}),
  (\ref{definition-Q-a}), (\ref{definition-ml-a}), and
  (\ref{definition-mj-a})] but:
\begin{gather}
  \sigma_1 = m_{s1}+\sigma\,,\quad
  \sigma_2 = m_{s2}-\sigma\,,\label{definition-sigma-aa}\displaybreak[0]\\
  m'_{s1} = m_{s1}+\sigma-\sigma'\,,\quad
  m'_{s2} = m_{s2}-\sigma+\sigma'\,,\label{definition-ms-aa}
\end{gather}

%%%%%%%%%%%%%%%%%%%%%%%%%%%%%%%%%%%%%%%%%%%%%%%%%%%%%%%%%%%%%%%%%%%%%%%%%%%%%%%%
\subsection{Diagram (b): RPA bubble summation}
%%%%%%%%%%%%%%%%%%%%%%%%%%%%%%%%%%%%%%%%%%%%%%%%%%%%%%%%%%%%%%%%%%%%%%%%%%%%%%%%
Let us now turn to diagram (b), which includes the RPA bubble
summation. In the present work, we will restrict ourselves to the
Landau approximation and keep only the lowest-order ($L=0$) Landau
parameters. Then the particle-hole interaction takes the form
$f+g\,\vek{\sigma}_1\cdot\vek{\sigma}_2$, which allows one to sum the
RPA bubble series separately in the $S=0$ and $S=1$ channels. The
resulting particle-hole interactions are then
\begin{equation}
f_{\text{RPA}}(k) = \frac{f_0}{1-f_0\Pi_0(k)}\,,\quad
g_{\text{RPA}}(k) = \frac{g_0}{1-g_0\Pi_0(k)}\,,
\label{RPA}
\end{equation}
where $f_0$ and $g_0$ are the Landau parameters for $S=0$ and $S=1$,
respectively, and
\begin{equation}
\Pi_0(k) = -2\int \frac{d^3p}{(2\pi)^3}\,
\frac{n(\vek{p}-\vek{k})-n(\vek{p})}
     {\epsilon(\vek{p})-\epsilon(\vek{p}-\vek{k})}
\end{equation}
is the static ($\omega\to 0$) limit of the usual Lindhard function
$\Pi_0(k,\omega)$.

It is convenient to introduce the dimensionless Landau parameters $F_0
= N_0 f_0$ and $G_0 = N_0 g_0$, where $N_0 = m^*k_F/\pi^2$ is the
density of states at the Fermi surface (including the neutron-matter
degeneracy factor of two), and the dimensionless Lindhard function
$\tilde{\Pi}_0 = \Pi_0/N_0$. Then, \Eq{RPA} can be rewritten as
\begin{equation}
f_{\text{RPA}}(k) = \frac{F_0/N_0}{1-F_0\tilde{\Pi}_0(k)}\,,\quad
g_{\text{RPA}}(k) = \frac{G_0/N_0}{1-G_0\tilde{\Pi}_0(k)}\,.
\end{equation}
At zero temperature, the Lindhard function can be given in closed form
\cite{FetterWalecka},
\begin{equation}
\tilde{\Pi}_0(k) =
  \frac{1}{2}\left[-1+\frac{1-\tilde{k}^2/4}{\tilde{k}}
    \ln\left|\frac{1-\tilde{k}/2}{1+\tilde{k}/2}\right|\right]\,,
\end{equation}
with $\tilde{k} = k/k_F$.

When computing diagram (b), we use again the trick explained in the
derivation of \Eq{eq:diagaa} and compute the $S=1$ contribution as
three times the $m_S=0$ term, for which $\sigma'_1 = \sigma_1$ and
$\sigma'_2 = \sigma_2$:
\begin{widetext}
\begin{multline}
V_{b}(q,q') = \frac{1}{4\pi}
  \sum_{\sigma\sigma'}
    \CG{\half}{\sigma}{\half}{-\sigma}{0}{0}
    \CG{\half}{\sigma'}{\half}{-\sigma'}{0}{0}
  \int \frac{d\Omega_{\vek{q}}}{4\pi}
  \int \frac{d\Omega_{\vek{q}'}}{4\pi}
  \int \frac{d^3p_1}{(2\pi)^3}\,
    \frac{n(\vek{p}_1-\vek{k})-n(\vek{p}_1)}
      {\epsilon(\vek{p}_1)-\epsilon(\vek{p}_1-\vek{k})}
  \int \frac{d^3p_2}{(2\pi)^3}\,
    \frac{n(\vek{p}_2-\vek{k})-n(\vek{p}_2)}
      {\epsilon(\vek{p}_2)-\epsilon(\vek{p}_2-\vek{k})}\\ \times
  \sum_{s_1 m_{s1}}\sum_{s_2 m_{s2}}
  \sum_{l_1l'_1m_{l1}} \sum_{l_2,l'_2,m_{l2}} \sum_{j_1j_2}
    \CG{\half}{\sigma_1}{\half}{-\sigma}{s_1}{m_{s1}}
    \CG{\half}{\sigma_1}{\half}{-\sigma'}{s_1}{m'_{s1}}
    \CG{\half}{\sigma}{\half}{\sigma_2}{s_2}{m_{s2}}
    \CG{\half}{\sigma'}{\half}{\sigma_2}{s_2}{m'_{s2}}
    \CG{l_1}{m_{l1}}{s_1}{m_{s1}}{j_1}{m_{j1}}
    \CG{l'_1}{m'_{l1}}{s_1}{m'_{s1}}{j_1}{m_{j1}}
    \CG{l_2}{m_{l2}}{s_2}{m_{s2}}{j_2}{m_{j2}}
    \CG{l'_2}{m'_{l2}}{s_2}{m'_{s2}}{j_2}{m_{j2}}\\ \times
    (4\pi)^4 i^{l'_1-l_1+l'_2-l_2} 
      Y^*_{l_1m_{l1}}(\Omega_{\vek{Q}_1}) 
      Y_{l'_1m'_{l1}}(\Omega_{\vek{Q}'_1})
      Y^*_{l_2m_{l2}}(\Omega_{\vek{Q}_2})
      Y_{l'_2m'_{l2}}(\Omega_{\vek{Q}'_2})
    \langle Q_1|\tilde{V}_{s_1l_1l'_1j_1}|Q'_1\rangle
    \langle Q_2|\tilde{V}_{s_2l_2l'_2j_2}|Q'_2\rangle \\ \times
    \left[f_{\text{RPA}}(k)
       +3\,(-1)^{1-m_{s1}-m_{s2}} g_{\text{RPA}}(k)\right]\,,
    \label{eq:diagb}
\end{multline}
\end{widetext}
with the same abbreviations $\vek{k}$, $\sigma_i$, $m'_{si}$,
$m'_{li}$, and $m_{ji}$ as before [Eqs. (\ref{definition-k-a}),
  (\ref{definition-sigma-aa}), (\ref{definition-ms-aa}),
  (\ref{definition-ml-a}), and (\ref{definition-mj-a})] but:
\begin{equation}
\begin{split}
  &\vek{Q}_1 = \frac{\vek{q}+\vek{p}_1}{2}\,,\quad
  \vek{Q}'_1 = \frac{\vek{q}'-\vek{k}+\vek{p}_1}{2}\,, \\
  &\vek{Q}_2 = \frac{\vek{q}+\vek{k}-\vek{p}_2}{2}\,,\quad
  \vek{Q}'_2 = \frac{\vek{q}'-\vek{p}_2}{2}\,,\label{definition-Q-b}
\end{split}
\end{equation}

%%%%%%%%%%%%%%%%%%%%%%%%%%%%%%%%%%%%%%%%%%%%%%%%%%%%%%%%%%%%%%%%%%%%%%%%%%%%%%%%
\section{Anti-screening due to the RPA}
\label{sec:RPA}
%%%%%%%%%%%%%%%%%%%%%%%%%%%%%%%%%%%%%%%%%%%%%%%%%%%%%%%%%%%%%%%%%%%%%%%%%%%%%%%%
\subsection{Parameters}
\label{sec:parameters}
%%%%%%%%%%%%%%%%%%%%%%%%%%%%%%%%%%%%%%%%%%%%%%%%%%%%%%%%%%%%%%%%%%%%%%%%%%%%%%%%
For the $nn$ interaction in the particle-particle channel, we use the
low-momentum interaction $\vlowk$ from \Ref{Bogner2007}, obtained from
the AV$_{18}$ interaction by a renormalization group evolution (using
a smooth Fermi-Dirac regulator with $\epsilon_{\text{FD}}=0.5$) to a
final cutoff of $\Lambda=2\fmi$.

For the purpose of comparing with \Ref{Shen2005}, we also perform
calculations with the Gogny force, using the D1 parameterization
\cite{Decharge1980} and the more recent D1N parameterization
\cite{Chappert2008}. For a comparison of the matrix elements of the
Gogny force with those of $\vlowk$, and the corresponding pairing gaps
without screening, see \Ref{Sedrakian2003}. The explicit expressions
for the partial-wave expansion of the Gogny force are given in
Appendix~\ref{app:Gogny}.

\begin{figure}
  \includegraphics[width=7cm]{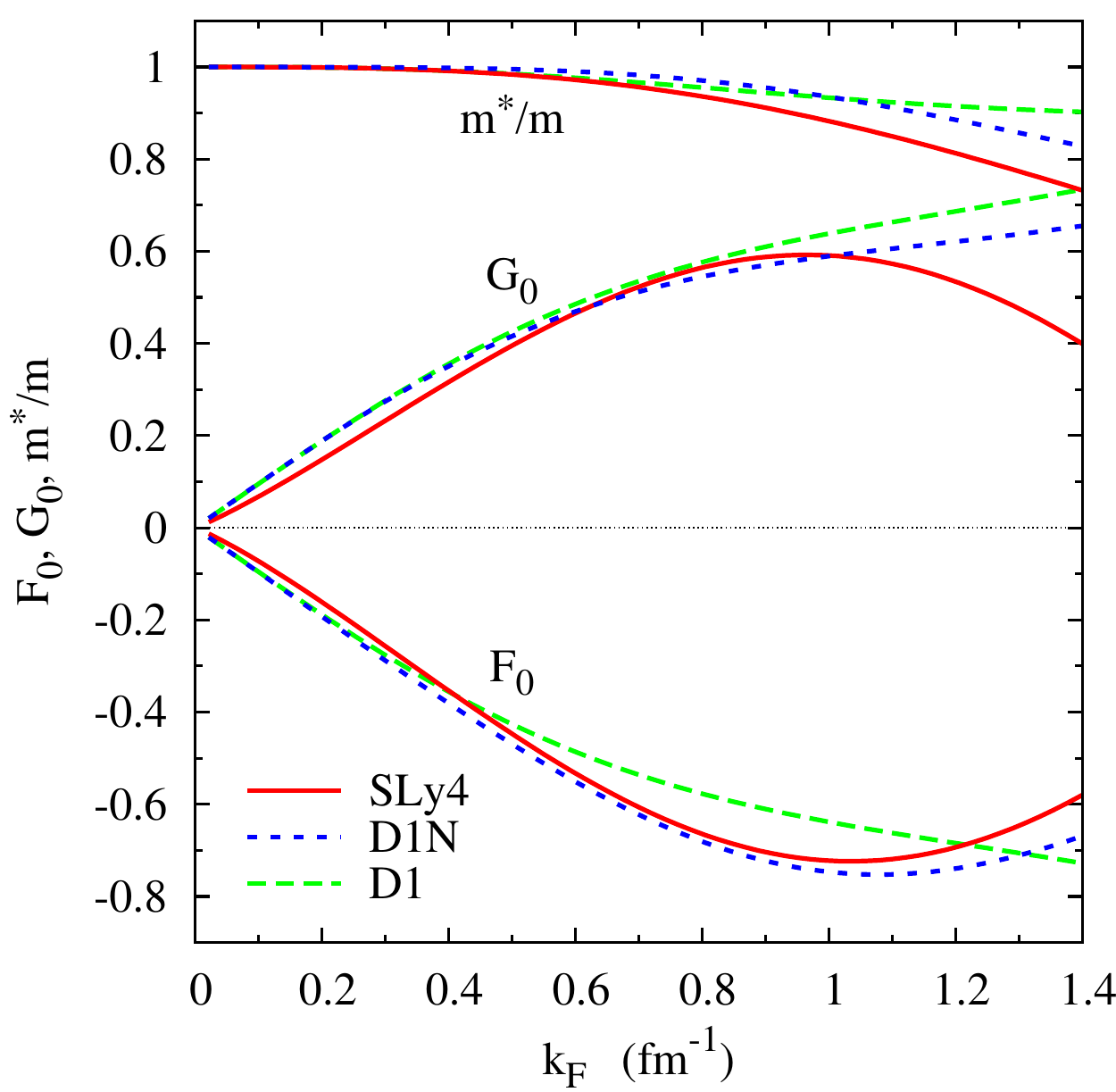}
\caption{\label{fig:Landau} Fermi-liquid parameters $m^*/m$,
    $G_0$, and $F_0$ used in the present work, obtained from different
    phenomenological effective interactions: Skyrme parameterization
    SLy4 (solid lines), and Gogny D1N (short dashes) and D1 (long
    dashes) parameterizations.}
\end{figure}

Concerning the Fermi-liquid parameters, we do not attempt to compute
them from the microscopic theory, but we take more phenomenological
results from the SLy4 parameterization of the Skyrme functional
\cite{Chabanat1997} or from the D1N parameterization of the Gogny
force \cite{Chappert2008}. The explicit formulas are given in
Appendix~\ref{app:Landau}, and the resulting Fermi-liquid parameters
$m^*/m$, $F_0$, and $G_0$ are shown in \Fig{fig:Landau}.
%%%%%%%%%%%%%%%%%%%%%%%%%%%%%%%%%%%%%%%%%%%%%%%%%%%%%%%%%%%%%%%%%%%%%%%%%%%%%%%%
%%%%%%%%%%%%%%%%%%%%%%%%%%%%%%%%%%%%%%%%%%%%%%%%%%%%%%%%%%%%%%%%%%%%%%%%%%%%%%%%
Since both the SLy4 and the D1N effective interactions have been
fitted to the neutron-matter equation of state, it is not surprising
that they give almost identical results for the Landau parameter $F_0$
at low densities. But also the $G_0$ values are quite close to each
other. Above $k_F\sim 1\fmi$, however, the Landau parameters of SLy4
are clearly smaller (in absolute value) than those of D1N. Note also
that SLy4 systematically yields a smaller effective mass $m^*$ than
D1N. For a comparison with \Ref{Shen2005}, we also used the D1
parameterization of the Gogny force \cite{Decharge1980}, the resulting
Fermi-liquid parameters are also shown in \Fig{fig:Landau}.

%%%%%%%%%%%%%%%%%%%%%%%%%%%%%%%%%%%%%%%%%%%%%%%%%%%%%%%%%%%%%%%%%%%%%%%%%%%%%%%%
\subsection{Induced interaction}
\label{sec:results:vind}
%%%%%%%%%%%%%%%%%%%%%%%%%%%%%%%%%%%%%%%%%%%%%%%%%%%%%%%%%%%%%%%%%%%%%%%%%%%%%%%%
In order to calculate the induced interaction in practice, we restrict
the partial-wave expansion in \Eq{eq:partialwaves} to some maximum
angular momentum, $j\leq j_{\text{max}}$. The multidimensional
integrals in \Eqs{eq:diaga} or (\ref{eq:diagaa}), and
(\ref{eq:diagb}) are computed using Monte-Carlo integration.  Data
files containing tables of the pairing interaction with and without
the induced interaction are provided in the supplemental material
\cite{suppl}.

\begin{figure}
\includegraphics[width=7cm]{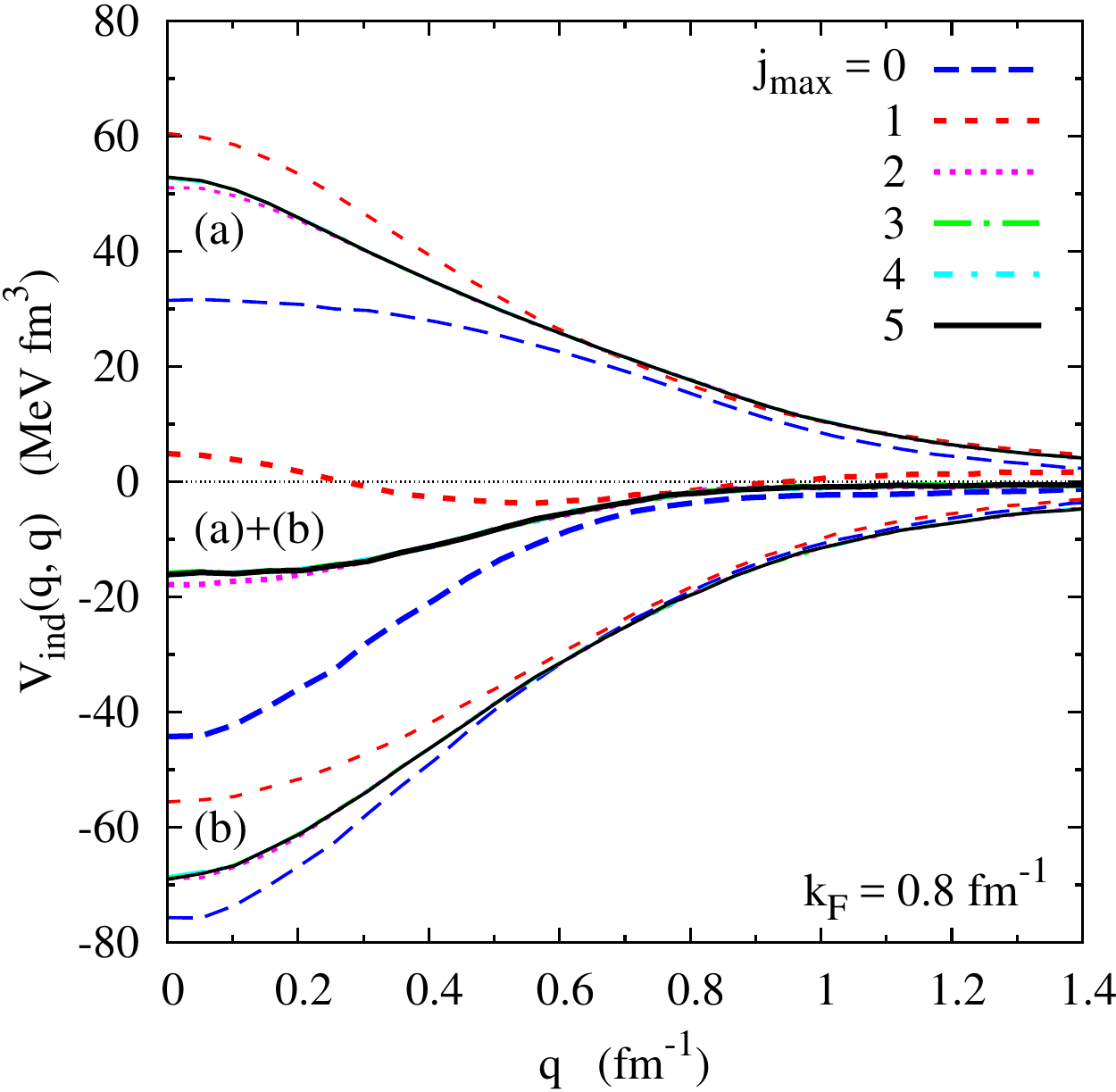}
\caption{\label{fig:convergence} Convergence of the induced
  interaction with respect to variation of the maximum angular
  momentum $j_{\text{max}}$ used in the partial wave expansion of the
  bare interaction. The figure shows the diagonal matrix elements for
  $k_F = 0.8\fmi$. Upper thin curves: results for diagram (a), lower
  thin curves: results for diagram (b), thick curves: sum (a)+(b). The
  bare interaction in this example is $\vlowk$, and the Fermi-liquid
  parameters are those of SLy4.}
\end{figure}

First, we have to check that convergence w.r.t. $j_{\text{max}}$ has
been reached. This is indeed the case for $j_{\text{max}} = 3$, as can
be seen in \Fig{fig:convergence}.
%%%%%%%%%%%%%%%%%%%%%%%%%%%%%%%%%%%%%%%%%%%%%%%%%%%%%%%%%%%%%%%%%%%%%%%%%%%%%%%%
%%%%%%%%%%%%%%%%%%%%%%%%%%%%%%%%%%%%%%%%%%%%%%%%%%%%%%%%%%%%%%%%%%%%%%%%%%%%%%%%
As one can see from this figure, for the example $k_F = 0.8\fmi$, the
net effect of the sum of diagrams (a) and (b) is attractive, i.e., the
strong repulsion generated by diagram (a) is more than compensated for by
the attractive diagram (b).

\begin{figure}
\includegraphics[width=7cm]{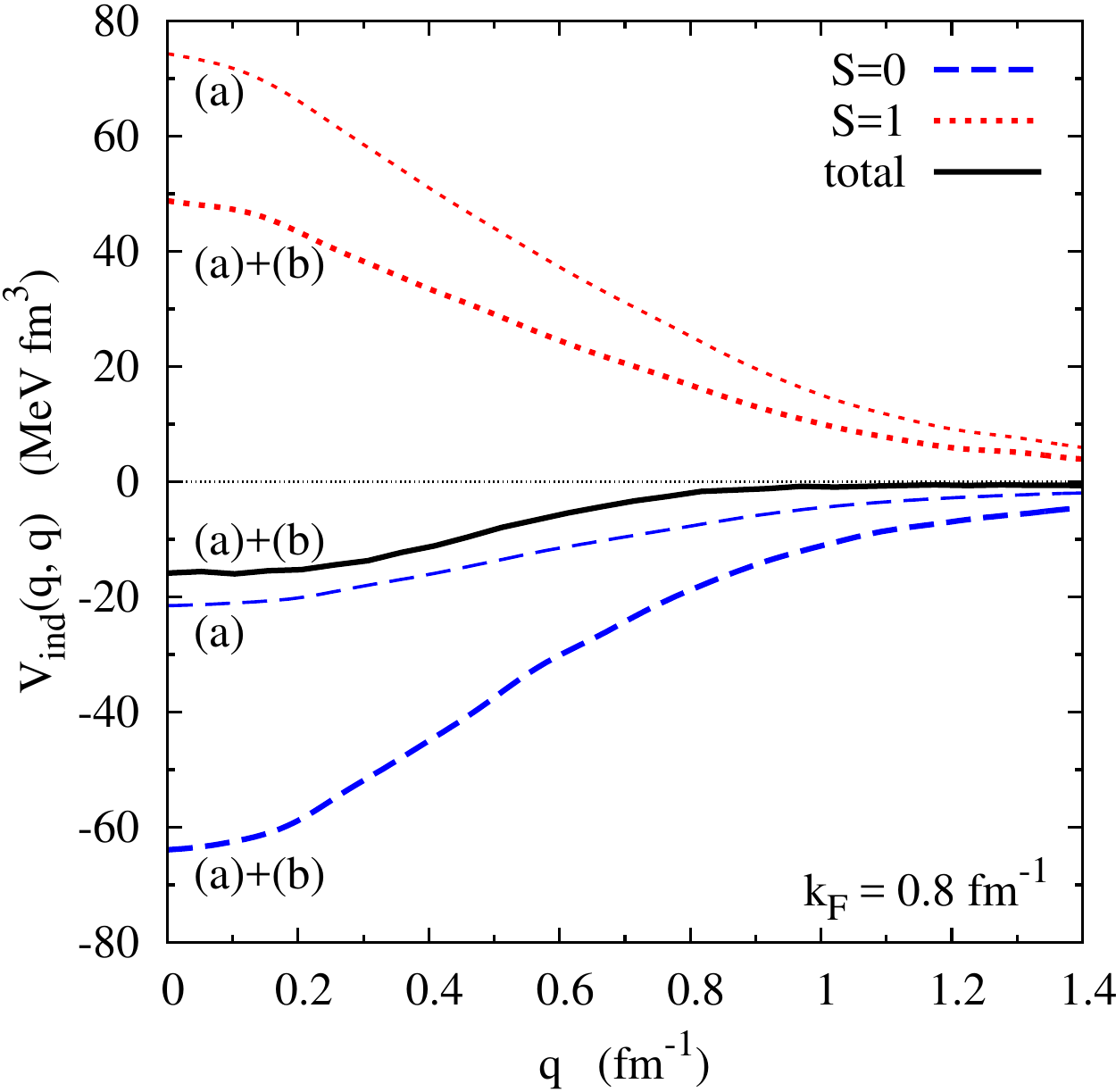}
\caption{\label{fig:ph-spins} Induced pairing interaction due to the
  exchange of $S=0$ (dashed lines) and $S=1$ (dotted lines)
  excitations. The thin lines represent the contributions of diagrams
  (a) only, and the thick lines are the sums of diagrams (a) and
  (b). The thick solid line is the sum of $S=0$ and $S=1$
  contributions. The parameters are the same as in
  \Fig{fig:convergence}.}
\end{figure}

This result is in contrast to previous studies \cite{Shen2005,Cao2006}
 where it was found that the contribution of diagram (b)
is attractive but not strong enough to compensate for the repulsion
generated by diagram (a). Let us therefore analyse our result in more
detail. It is known that the exchange of $S=0$ excitations (density
fluctuations) is attractive and that of $S=1$ excitations
(spin-density fluctuations) is repulsive
\cite{Heiselberg2000,Cao2006}. This is also the case in our
calculation, as shown in \Fig{fig:ph-spins}, again for the example
$k_F=0.8\fmi$.
%%%%%%%%%%%%%%%%%%%%%%%%%%%%%%%%%%%%%%%%%%%%%%%%%%%%%%%%%%%%%%%%%%%%%%%%%%%%%%%%
%%%%%%%%%%%%%%%%%%%%%%%%%%%%%%%%%%%%%%%%%%%%%%%%%%%%%%%%%%%%%%%%%%%%%%%%%%%%%%%%
If there was only the single bubble exchange [diagram (a)], the
repulsive contribution of $S=1$ excitations would be three to four
times larger than the attractive one of $S=0$ excitations. However,
the inclusion of the RPA [diagram (b)] acts differently in the cases
$S=0$ and $S=1$ because the Landau parameters have opposite signs. In
the $S=0$ case, since $F_0 < 0$, the effect of diagram (a) is
enhanced, while in the $S=1$ case, since $G_0 > 0$, the effect of
diagram (a) is reduced. Therefore, with the inclusion of the RPA, the
attraction due to the exchange of density waves can finally win
against the repulsive effect of the spin-density waves.

%%%%%%%%%%%%%%%%%%%%%%%%%%%%%%%%%%%%%%%%%%%%%%%%%%%%%%%%%%%%%%%%%%%%%%%%%%%%%%%%
\subsection{Critical temperature}
\label{sec:tc}
\begin{figure}
\includegraphics[width=7cm]{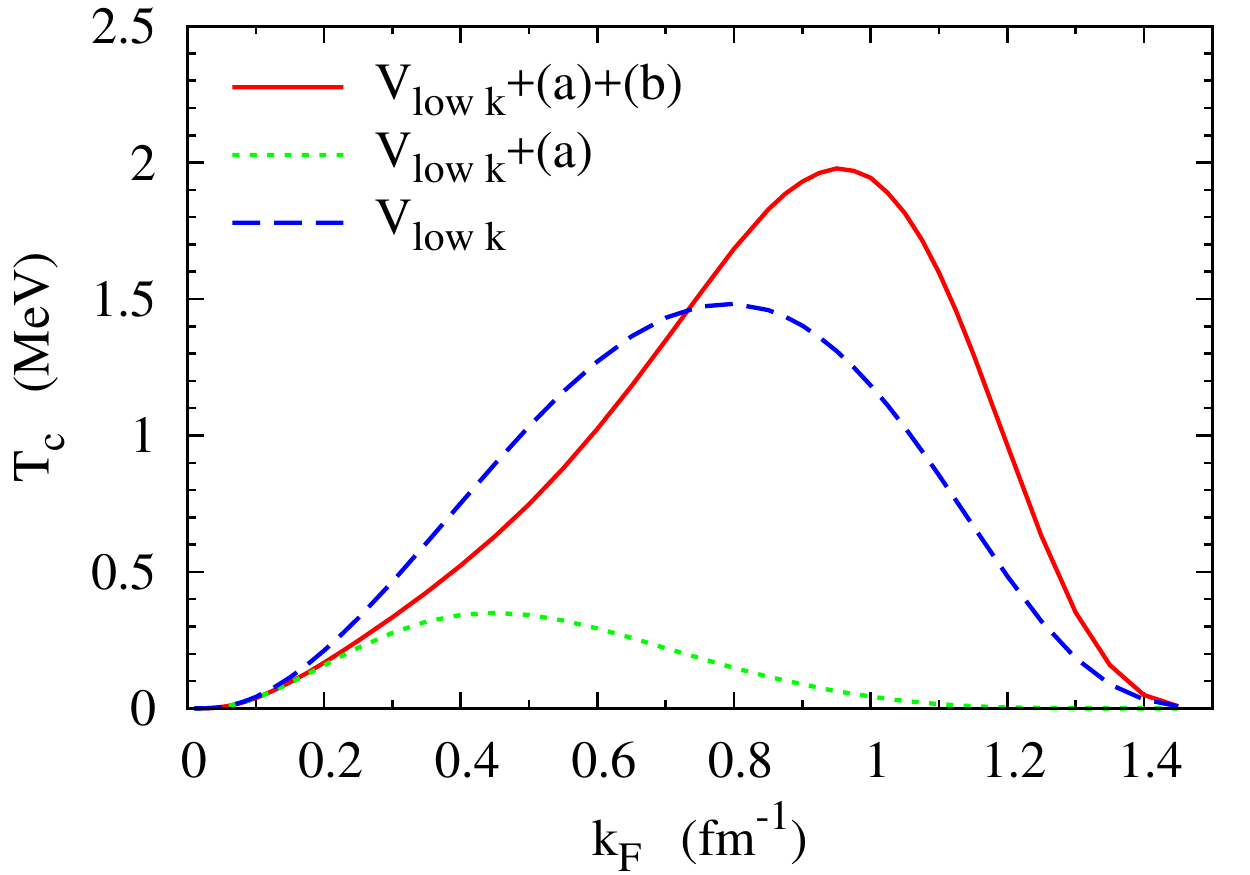}
\caption{\label{fig:Tc_vlowk} Critical temperature $T_c$ as a function
  of the Fermi momentum $k_F$, obtained with the $\vlowk$ interaction
  (Fermi-liquid parameters from the Skyrme force SLy4). Dashes: result
  obtained using only the bare interaction; dots: result obtained
  including diagram (a); solid line: full result including also
  diagram (b).}
\end{figure}

%%%%%%%%%%%%%%%%%%%%%%%%%%%%%%%%%%%%%%%%%%%%%%%%%%%%%%%%%%%%%%%%%%%%%%%%%%%%%%%%
We can now use the induced interaction $\Vind = V_a+V_b$ and
replace the bare interaction $V_0$ in the gap equation (\ref{gapeq})
by $V_0+\Vind$. The resulting critical temperature $T_c$ as a
function of the Fermi momentum $k_F$ is shown in \Fig{fig:Tc_vlowk}.
%%%%%%%%%%%%%%%%%%%%%%%%%%%%%%%%%%%%%%%%%%%%%%%%%%%%%%%%%%%%%%%%%%%%%%%%%%%%%%%%
%%%%%%%%%%%%%%%%%%%%%%%%%%%%%%%%%%%%%%%%%%%%%%%%%%%%%%%%%%%%%%%%%%%%%%%%%%%%%%%%
A sample of the results is also listed in Table~\ref{tab:Tc_vlowk}.
%%%%%%%%%%%%%%%%%%%%%%%%%%%%%%%%%%%%%%%%%%%%%%%%%%%%%%%%%%%%%%%%%%%%%%%%%%%%%%%%
\begin{table}
\caption{\label{tab:Tc_vlowk} Critical temperature as a function
    of the Fermi momentum $k_F$, obtained with $\vlowk$ interactions
    and Fermi-liquid parameters from SLy4. $T_c^{\text{(bare)}}$ is
    obtained with the bare interaction, while
    $T_c^{\text{(screened)}}$ includes the effect of $\Vind =
    V_a+V_b$. The columns marked $\Lambda=2$ fm$^{-1}$ correspond to
    the parameters given in \Sec{sec:parameters}, while for the
    columns marked $\Lambda=2.5\,k_F$, a $\vlowk$ interaction with a
    density dependent cutoff and a different regulator was used, see
    \Sec{sec:ddcutoff}.}
  \begin{ruledtabular}
    \newcommand{\head}[1]{\multicolumn{1}{c}{#1}}
    \begin{tabular}{D{.}{.}{1.2}*{4}{D{.}{.}{1.4}}}
      \head{}& \multicolumn{2}{c}{$\Lambda = 2$ fm$^{-1}$} &
        \multicolumn{2}{c}{$\Lambda = 2.5\,k_F$}\\ \hline
      \head{$k_F$} & \head{$T_c^{\text{(bare)}}$} & \head{$T_c^{\text{(screened)}}$} &
        \head{$T_c^{\text{(bare)}}$} & \head{$T_c^{\text{(screened)}}$}\\
      \head{(fm$^{-1}$)} & \head{(MeV)} & \head{(MeV)} &
        \head{(MeV)} & \head{(MeV)} \\ \hline
      0.08 & 0.0230 & 0.0211 & 0.0221 & 0.0135 \\
      0.2  & 0.212  & 0.167  & 0.206  & 0.128  \\
      0.4  & 0.752  & 0.523  & 0.743  & 0.488  \\
      0.6  & 1.27   & 1.02   & 1.27   & 1.03   \\
      0.8  & 1.48   & 1.68   & 1.48   & 1.70   \\
      1.0  & 1.18   & 1.94   & 1.18   & 1.90   \\
      1.2  & 0.485  & 0.964  & 0.474  & 0.789  \\
      1.3  & 0.184  & 0.352\\
      1.4  & 0.0323 & 0.0489 
    \end{tabular}
  \end{ruledtabular}
\end{table}
%%%%%%%%%%%%%%%%%%%%%%%%%%%%%%%%%%%%%%%%%%%%%%%%%%%%%%%%%%%%%%%%%%%%%%%%%%%%%%%%
The corresponding pairing gaps $\Delta(k_F)$ at $T=0$ can be obtained,
to a very good approximation, by multiplying $T_c$ with 1.76. The
dashed line represents the result obtained with the bare interaction
$\vlowk$. The maximum critical temperature is reached at $k_F\approx
0.8\fmi$. When one includes the induced interaction due to diagram (a)
alone, pairing is very strongly suppressed, as shown by the dotted
line. Finally, when including diagrams (a) and (b), one finds that the
critical temperature is lowered at low density, but increased at high
density. The change from screening to anti-screening is at $k_F\approx
0.73\fmi$, consistent with our results discussed in
\Sec{sec:results:vind} where we found that at $0.8\fmi$ the attractive
effect of $S=0$ excitations is stronger than the repulsive effect of
$S=1$ excitations. Whether the net effect of the induced interaction
is attractive (i.e., anti-screening) or repulsive (i.e. screening),
depends of course on the density and on the values of the Landau
parameters. With decreasing density, the RPA bubbles of diagram (b)
become less important and therefore the repulsive effect of diagram
(a) wins. This explains why, at very low density, the full result and
the result obtained with only diagram (a) become equal, as one can
also see in \Fig{fig:Tc_vlowk}.

\begin{figure}
\includegraphics[width=7cm]{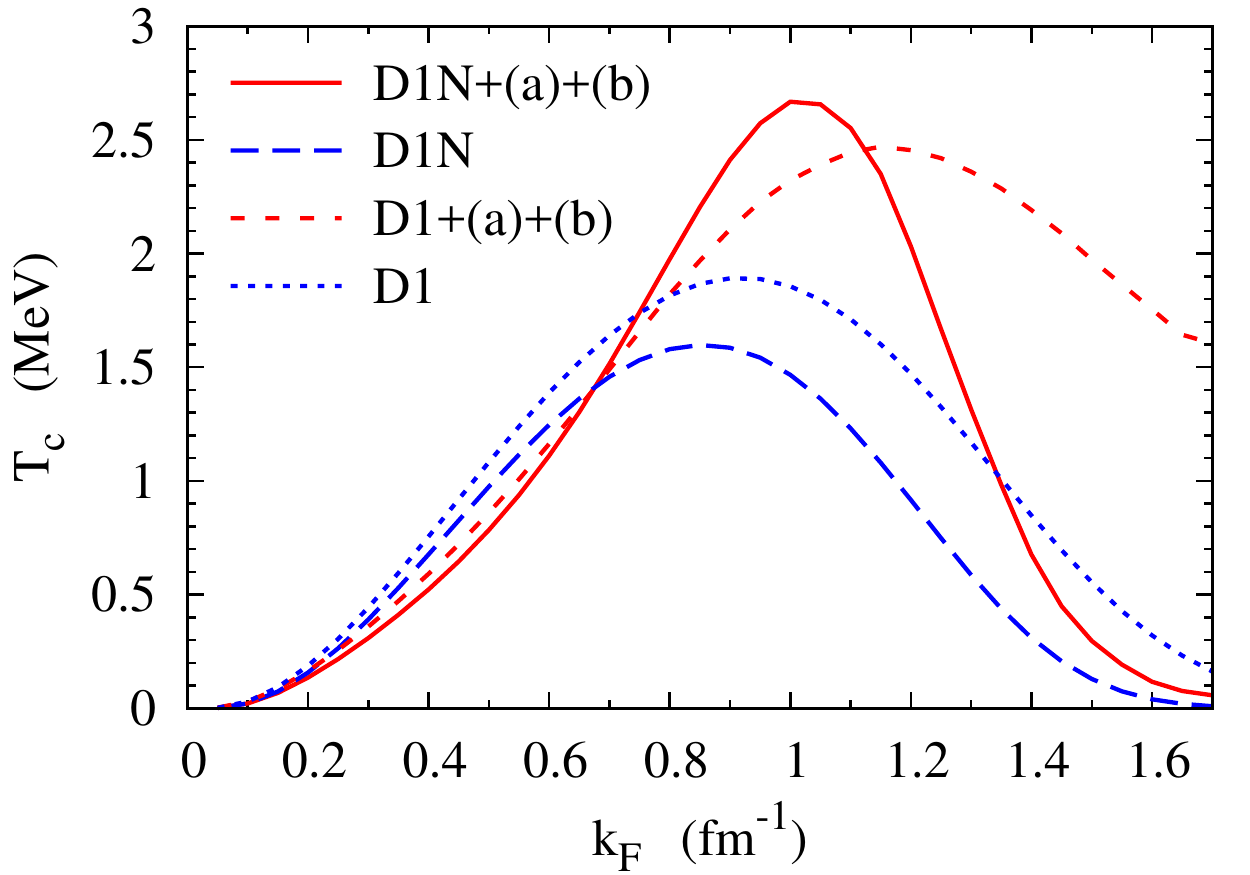}
\caption{\label{fig:Tc_D1N} Same as \Fig{fig:Tc_vlowk}, but here
  the Gogny D1N and D1 interactions are used in the particle-particle channel
  and for the Fermi-liquid parameters.}
\end{figure}

To check how sensitive our results are to the details of the model, we
repeated the calculation with the D1N and D1 Gogny forces. In these cases, the
same interaction is used for the bare pairing force, for the vertices
entering the induced interaction diagrams (a) and (b), and for the
Fermi-liquid parameters. The results are shown in
\Fig{fig:Tc_D1N}.
%%%%%%%%%%%%%%%%%%%%%%%%%%%%%%%%%%%%%%%%%%%%%%%%%%%%%%%%%%%%%%%%%%%%%%%%%%%%%%%%
%%%%%%%%%%%%%%%%%%%%%%%%%%%%%%%%%%%%%%%%%%%%%%%%%%%%%%%%%%%%%%%%%%%%%%%%%%%%%%%%
Of course, since the $^1S_0$ matrix elements of the different
interactions are not the same, there is already some difference at the
level of the bare interaction \cite{Sedrakian2003}: the maximum is
slightly shifted and the gap survives up to higher density. However,
the effect of the induced interaction is qualitatively the same as in
\Fig{fig:Tc_vlowk}, i.e., the gap is reduced at low density and
increased at high density. The change from screening to anti-screening
happens at about the same density as with $\vlowk$ (with Fermi-liquid 
parameters from SLy4) in \Fig{fig:Tc_vlowk}, and
compared to the $\vlowk$ results the anti-screening effect at high
density is even stronger with both the D1N and the D1 Gogny
interactions.

%%%%%%%%%%%%%%%%%%%%%%%%%%%%%%%%%%%%%%%%%%%%%%%%%%%%%%%%%%%%%%%%%%%%%%%%%%%%%%%%
\section{The low-density limit}
\label{sec:low-density-limit}
%%%%%%%%%%%%%%%%%%%%%%%%%%%%%%%%%%%%%%%%%%%%%%%%%%%%%%%%%%%%%%%%%%%%%%%%%%%%%%%%
As one sees from \Fig{fig:Tc_vlowk}, with the $\vlowk$ interaction
with a fixed cutoff of $2\fmi$, screening gets weak at low density and
finally at $k_F\lesssim 0.1\fmi$ one recovers the BCS result. 
However, at $k_F\ll 1/|a|$, the GMB result
should be valid, predicting a reduction of $T_c$ by a factor of
$(4e)^{-1/3} \approx 0.45$. Therefore, let us study the low-density
limit in more detail.

%%%%%%%%%%%%%%%%%%%%%%%%%%%%%%%%%%%%%%%%%%%%%%%%%%%%%%%%%%%%%%%%%%%%%%%%%%%%%%%
\subsection{Failure of the weak coupling formula}
\label{sec:weak-coupling-fails}
%%%%%%%%%%%%%%%%%%%%%%%%%%%%%%%%%%%%%%%%%%%%%%%%%%%%%%%%%%%%%%%%%%%%%%%%%%%%%%%
As we have seen, the contribution of diagram (b) becomes negligible at
low density. Concerning diagram (a), it seems natural to concentrate
on matrix elements $V_{a}(q,q')$ with $q,q'\simeq k_F$. If $k_F$
becomes small, this means that also $q$ and $q'$ and hence all the
momenta $Q_1$ etc. that appear in \Eq{eq:diaga} become
small. Therefore, we can replace
\begin{equation}
  \langle Q_i|\tilde{V}_{s_il_il_i'j_i}|Q_i'\rangle
    \xrightarrow{q,q',k_F\to 0}
  2 V_0(0,0)\, \delta_{s_i 0}\, \delta_{l_i0}\, \delta_{l_i'0}
  \,\delta_{j_i0}
\end{equation}
(the factor of two accounts for the antisymmetrization of $\tilde{V}$),
and \Eq{eq:diaga} simplifies tremendously to
\begin{equation}
  V_{a}(q,q') \approx -2\pi N_0
  |V_0(0,0)|^2 \langle\tilde{\Pi}_0\rangle\,.
  \label{eq:Va-approx}
\end{equation}
In this expression, we have used the angle-averaged Lindhard function
\begin{equation}
  \langle \tilde{\Pi}_0\rangle = \frac{1}{2}\int_{-1}^1 \!d\cos\theta\,
  \tilde{\Pi}_0\big(\sqrt{q^2+q^{\prime\,2}-2qq'\cos\theta}\big)\,,
  \label{eq:Piav}
\end{equation}
see appendix \ref{app:Piav}. In particular, we get
\begin{equation}
  V_{a}(k_F,k_F) \approx 2\pi N_0 |V_0(0,0)|^2 \tfrac{1}{3}\ln 4e\,.
  \label{eq:Va-kF-approx}
\end{equation}

Following well-known weak-coupling arguments \cite{FetterWalecka}, the
gap and critical temperature should be proportional to $e^{1/[2\pi
    N_0V_0(k_F,k_F)]}$. If we replace $V_0$ by
$V_0+V_{a}$ in the approximation given in \Eq{eq:Va-kF-approx}, we find that
the gap and the critical temperature should indeed be reduced by the
factor $(4e)^{-1/3}$, in contradiction to our numerical results which
show that at low density $T_c$ is not modified at all by
screening. Obviously the weak-coupling formula does not apply in the
present case, although we are clearly in a weak coupling situation
since $T_c \ll \epsilon_F$. Note that there are a couple of cases in
nuclear physics where the weak coupling formula is known to fail
\cite{Clark2013}.

When using the weak coupling formula, one assumes that the kernel
$\Kernel(k,q)$ given in \Eq{eq:kernel} is sharply peaked at $q = k_F$ and
that this peak gives the dominant contribution to the integral in the
gap equation. However, we will show that the contribution of the peak
is \textit{not} dominant at low density, and this is the reason
why the weak coupling formula fails in this case.

\begin{figure}
\includegraphics[width=7cm]{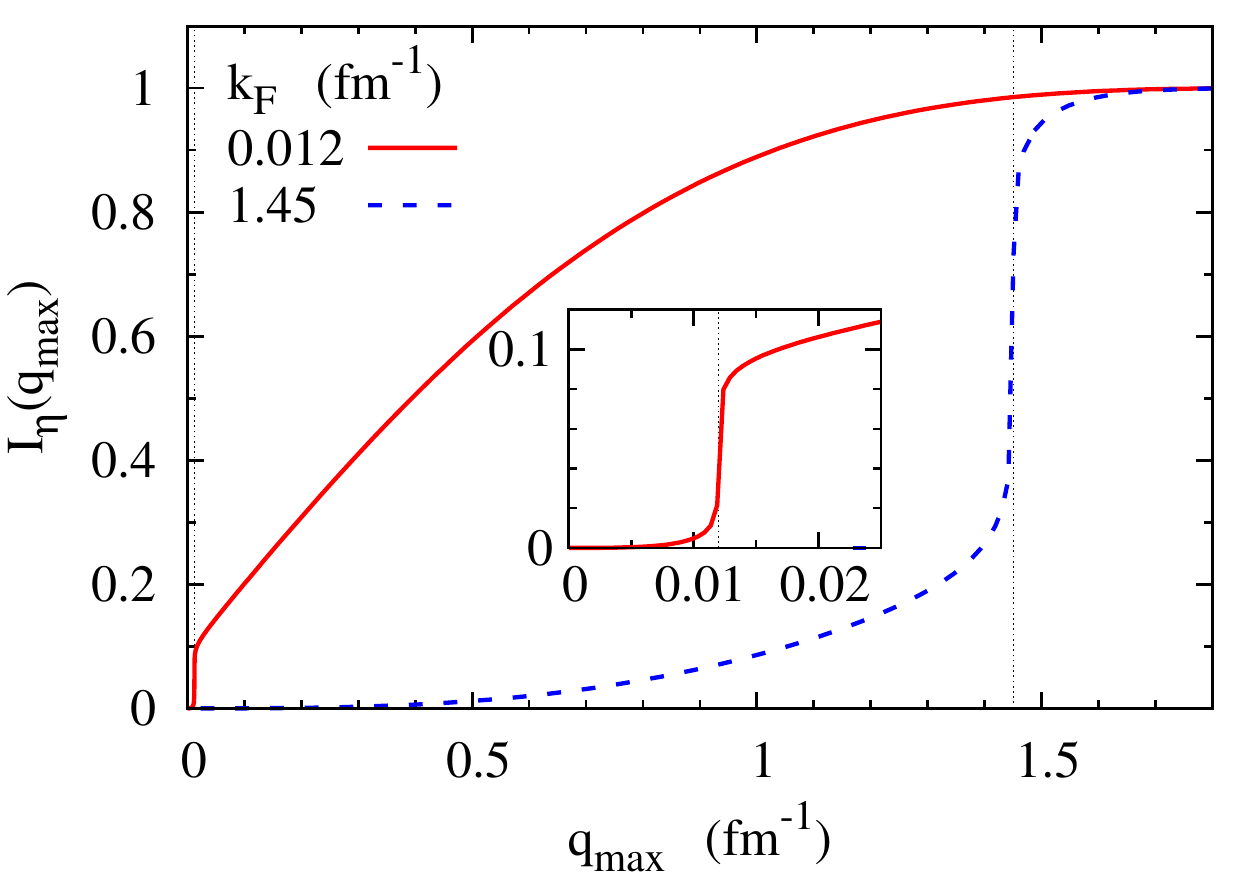}
\caption{\label{fig:Ieta} Measure of the contribution of different
  momenta to the gap equation as defined in \Eq{eq:Ieta}, for two
  different densities ($k_F = 0.012\fmi$ (solid line) and $1.45\fmi$
  (dashes), indicated by the thin vertical lines). The integrals were
  calculated with the $\vlowk$ interaction $m^*$ from the Skyrme force SLy4) at the
  respective critical temperatures.}
\end{figure}

Remember that the critical temperature is given by the temperature
where the largest eigenvalue $\eta$ of the kernel $\Kernel(k,q)$ given
in \Eq{eq:kernel} is equal to unity. The corresponding eigenvector
$|\phi\rangle$ can be found by numerical diagonalization, its
representation in momentum space, $\phi(q) = \langle q|\phi\rangle$,
is a smooth function of $q$ which has approximately the shape of
$V_0(q,k_F)$. If we normalize the eigenvector to
$\langle\phi|\phi\rangle = (2/\pi) \int\! dq\, q^2 |\phi(q)|^2 = 1$,
we can write the eigenvalue $\eta$ as
\begin{equation}
  \eta = \langle\phi|\Kernel|\phi\rangle = \frac{4}{\pi^2}\int\! dq\, q^2
  \int\! dk\, k^2 \phi(k) \Kernel(k,q) \phi(q)\,.
\end{equation}
To measure the importance of the peak
of the kernel at $q=k_F$, we can look at this integral as a function
of its upper limit $\qmax$,
\begin{multline}
  I_\eta(\qmax)
  = -\frac{4}{\pi^2} \int_0^{\qmax} dq\, q^2 \phi(q)
  \frac{\tanh\big(\frac{\xi(q)}{2T}\big)}{2\xi(q)}\\
  \times \int_0^\infty dk\, k^2 \phi(k) V(k,q)\,.
  \label{eq:Ieta}
\end{multline}
At $T=T_c$, we know that $I_\eta\to 1$ for $\qmax\to\infty$ since
$\eta=1$. For the weak coupling formula to be valid, the main
contribution to the integral should come from $q\approx k_F$, i.e.,
$I_\eta$ should be close to the step function $\theta(\qmax-k_F)$. In
\Fig{fig:Ieta}
%%%%%%%%%%%%%%%%%%%%%%%%%%%%%%%%%%%%%%%%%%%%%%%%%%%%%%%%%%%%%%%%%%%%%%%%%%%%%%%%
%%%%%%%%%%%%%%%%%%%%%%%%%%%%%%%%%%%%%%%%%%%%%%%%%%%%%%%%%%%%%%%%%%%%%%%%%%%%%%%%
we show the behavior of $I_\eta$ for two cases, $k_F = 1.45\fmi$
(dashes) and $k_F = 0.012\fmi$ (solid line). In both cases, we are in
the weak-coupling limit, in the sense that $T_c/E_F$ is very small (of
the order of $10^{-4}$). In the case $k_F = 1.45\fmi$, we see that
about 80\% of the integral come from momenta close to $k_F$, so that
in this case $T_c$ is indeed determined to a large extent by
$V(k_F,k_F)$. But in the low-density case, $k_F = 0.012\fmi$, the
situation is completely different. Although there is again a sharp
rise of $I_\eta$ at $q\approx k_F$ (visible in the zoom), its
contribution to the total integral is less than 10\%. The largest
contribution to the integral comes from momenta that are considerably
larger than $k_F$.

\begin{figure}
\includegraphics[width=7cm]{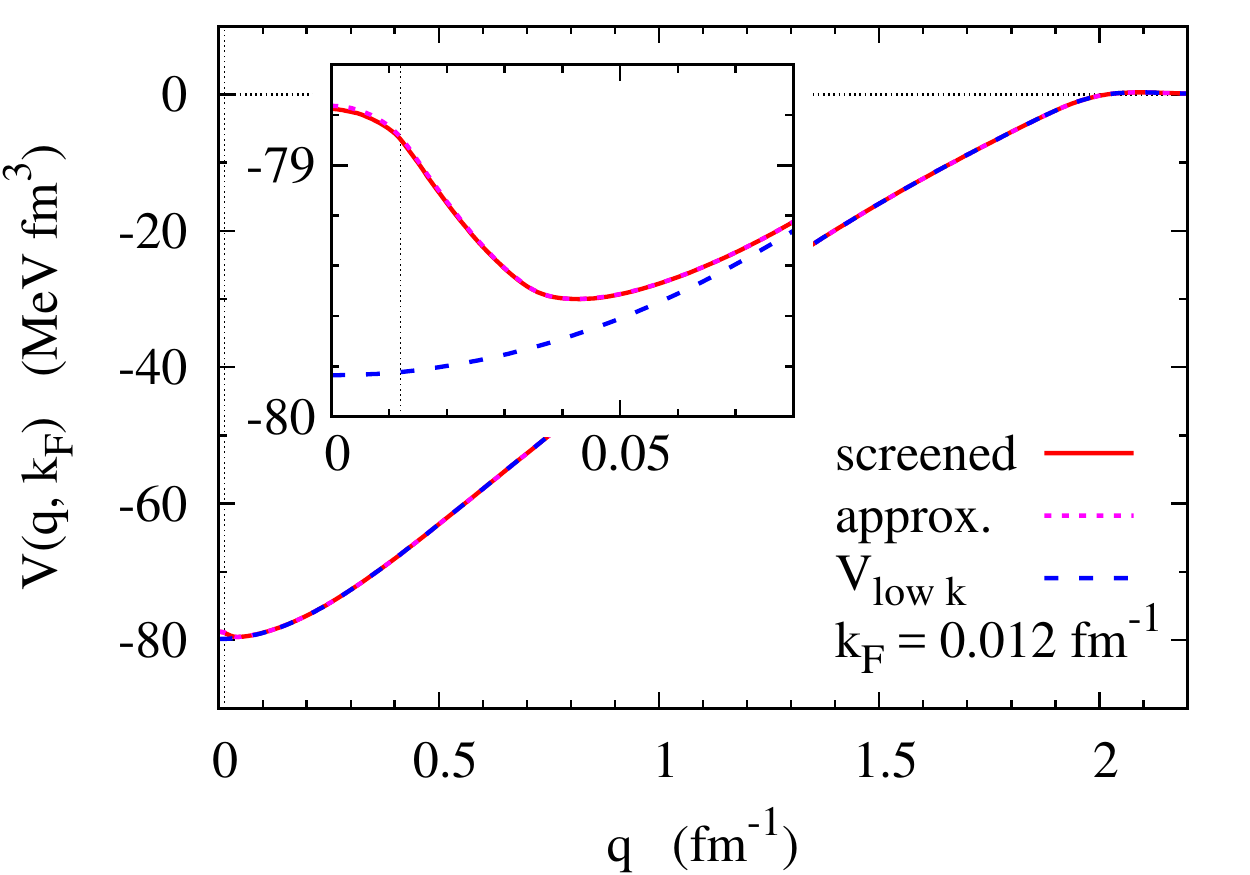}
\caption{\label{fig:v-low-dens-vlowk} Matrix elements $V(q,k_F)$ of
  the bare ($\vlowk$, dashes) and of the screened (solid line)
  interaction for $k_F = 0.012\fmi$. For comparison, we display also
  the screened interaction obtained with the analytical approximation
  \Eq{eq:Va-approx} (dots, almost indistinguishable from the solid
  line). The screening correction is so tiny that it is almost
  invisible on the big graph, see the inset for a zoom. The thin
  vertical line indicates $q=k_F$.}
\end{figure}

Let us now look at the matrix elements $V(q,k_F)$ for $k_F =
0.012\fmi$ with and without screening, which are displayed in
\Fig{fig:v-low-dens-vlowk}.
%%%%%%%%%%%%%%%%%%%%%%%%%%%%%%%%%%%%%%%%%%%%%%%%%%%%%%%%%%%%%%%%%%%%%%%%%%%%%%%%
%%%%%%%%%%%%%%%%%%%%%%%%%%%%%%%%%%%%%%%%%%%%%%%%%%%%%%%%%%%%%%%%%%%%%%%%%%%%%%%%
The screening correction is limited to the tiny region $q\lesssim
0.05\fmi \sim 4 k_F$, because of the strong momentum dependence of
the angle-averaged Lindhard function. But as we have seen before, this
small region contributes only about 10\% to the integral in the gap
equation, and therefore the screening correction has practically no
effect on the gap or $T_c$.

The observation that the screening effect disappears at low density is
not a singular feature of our calculation, but it can also be found in
the existing literature \cite{Cao2006}. However, as we will discuss
below, there are other problems with the low-density limit. Taking
these into account, we will eventually retrieve the GMB result.

%%%%%%%%%%%%%%%%%%%%%%%%%%%%%%%%%%%%%%%%%%%%%%%%%%%%%%%%%%%%%%%%%%%%%%%%%%%%%%%
\subsection{Failure of perturbation theory and density-dependent cutoff}
\label{sec:ddcutoff}
%%%%%%%%%%%%%%%%%%%%%%%%%%%%%%%%%%%%%%%%%%%%%%%%%%%%%%%%%%%%%%%%%%%%%%%%%%%%%%%
When calculating diagrams (a) and (b), we use the bare interaction $V$
perturbatively to describe the vertex coupling the particles to the
particle-hole excitations. Since we are using renormalized
interactions whose matrix elements decrease rapidly with increasing
relative momenta $Q_i$ and $Q_i^\prime$, which are typically of the
order of $k_F$, this may be a good approximation at higher densities.
However, for small $Q_i$ and $Q_i^\prime$, as they appear at low
densities, we know from the large value of the $nn$ scattering
length $a$ that the perturbative treatment must
fail~\cite{Bogner-nm-2005, Bogner-convergence-2006}.

\begin{figure}
\includegraphics[scale=0.9]{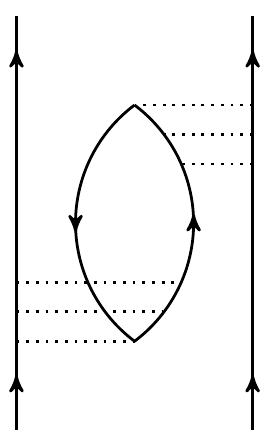}
\caption{\label{fig:diagram-a-resummed} 
Higher-order ladder diagrams in the 3p1h vertices
 which are not included in the present work.}
\end{figure}

When looking at the historical work by GMB \cite{Gorkov1961}, one
observes that they compute the correction in a different way. Namely,
instead of using the potential $V$ in the dashed interaction vertices of 
diagram (a), they use $a/m$. This amounts to including, at least
approximately, the resummation of ladder diagrams as shown in
\Fig{fig:diagram-a-resummed}.
%%%%%%%%%%%%%%%%%%%%%%%%%%%%%%%%%%%%%%%%%%%%%%%%%%%%%%%%%%%%%%%%%%%%%%%%%%%%%%%%
%%%%%%%%%%%%%%%%%%%%%%%%%%%%%%%%%%%%%%%%%%%%%%%%%%%%%%%%%%%%%%%%%%%%%%%%%%%%%%%%

In contrast to the Gogny interaction, the renormalization-group
evolved $\vlowk$ interaction gives us the additional freedom to change
the cutoff $\Lambda$. On the one hand, by lowering the cutoff, the
interaction gets obviously ``more perturbative''. In this sense, it is
tempting to lower the cutoff as much as possible. In fact, for
$q,q'<\Lambda$ and $\Lambda\to 0$, the matrix elements get more and
more attractive and flow towards the constant $a/m$ as
\begin{equation}
  V_0(q,q')\approx \Big(\frac{m}{a}-\frac{2m\Lambda}{\pi}\Big)^{-1}\,.
\label{eq:flow}
\end{equation}
This means that the contribution of higher-order ladder diagrams gets
progressively included, via the renormalization group flow, in the
two-body matrix elements, while the loop integrals become
  suppressed, and as a result, it should be possible to work with a
Born approximation to the $\Tmatrix$ matrix at low cutoffs. On the other hand, one of course
must not lower the cutoff below the relevant momentum scale of the
order of $k_F$.

The cutoff dependence of the gap (without screening corrections) was
investigated in \Ref{Schwenk2007}. Numerically, we obtain cutoff
independent results for $T_c$ at the BCS level in the whole range of
densities for $\Lambda\gtrsim 2.5\, k_F$, if we use an exponential
regulator of the form $\exp(-(k^2/\Lambda^2)^{n_{\text{exp}}})$ with
  $n_{\text{exp}}=5$. (With the Fermi-Dirac regulator and with  
  $\epsilon_{\text{FD}} = 0.5 \fmi$ 
  that we used before we would need somewhat larger cutoffs.)

So, let us see what we find when we choose instead of a constant
cutoff $\Lambda=2\fmi$ the lowest possible cutoff for each value of
$k_F$, i.e., $\Lambda = 2.5\, k_F$.

\begin{figure}
\includegraphics[width=7cm]{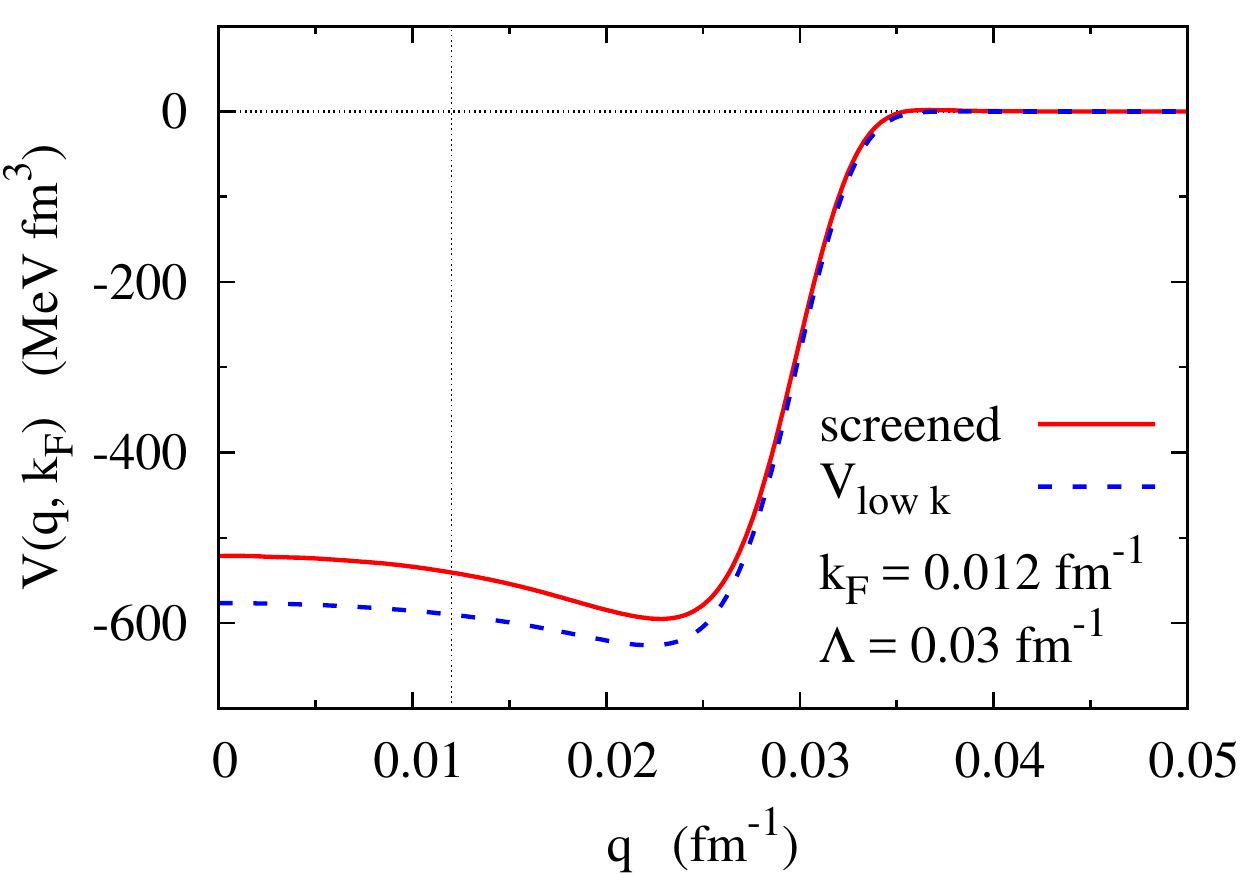}
\caption{\label{fig:v-low-dens-vlvar} Same as
  \Fig{fig:v-low-dens-vlowk} but now calculated with an interaction
  $\vlowk$ evolved to a much lower cutoff $\Lambda = 0.03\fmi = 2.5\,
  k_F$ (and with an exponential instead of Fermi-Dirac regulator, see text).}
\end{figure}

As an example, let us consider as in \Fig{fig:v-low-dens-vlowk} the
case $k_F = 0.012\fmi$. If we evolve the cutoff to the lowest possible
value for this $k_F$, i.e., to $\Lambda = 2.5\, k_F = 0.03\fmi$, we
obtain the matrix elements $V(q,k_F)$ shown in
\Fig{fig:v-low-dens-vlvar}.
%%%%%%%%%%%%%%%%%%%%%%%%%%%%%%%%%%%%%%%%%%%%%%%%%%%%%%%%%%%%%%%%%%%%%%%%%%%%%%%%
%%%%%%%%%%%%%%%%%%%%%%%%%%%%%%%%%%%%%%%%%%%%%%%%%%%%%%%%%%%%%%%%%%%%%%%%%%%%%%%%
As in \Fig{fig:v-low-dens-vlowk}, the dashed line represents $\vlowk$
without screening and the solid line has screening included. The most
obvious difference between \Figs{fig:v-low-dens-vlowk} and
\ref{fig:v-low-dens-vlvar} is that, when the cutoff is lowered, the
$\vlowk$ matrix elements (dashed lines) get more attractive,
cf. \Eq{eq:flow}. However, the renormalization group flow does not
only ensure that the low-energy scattering in free space remains
unchanged, but also the gap and $T_c$ at the BCS level (i.e., without
screening) remain the same, as mentioned above. But the results with
screening change. Now, the modification of the interaction due to
screening (difference between the solid and the dashed lines in
  \Fig{fig:v-low-dens-vlvar}) extends over the whole momentum range
up to $\sim \Lambda$, and therefore the screening will reduce $T_c$,
contrary to what happened in the case $\Lambda=2\fmi$.

\begin{figure}
\includegraphics[width=7cm]{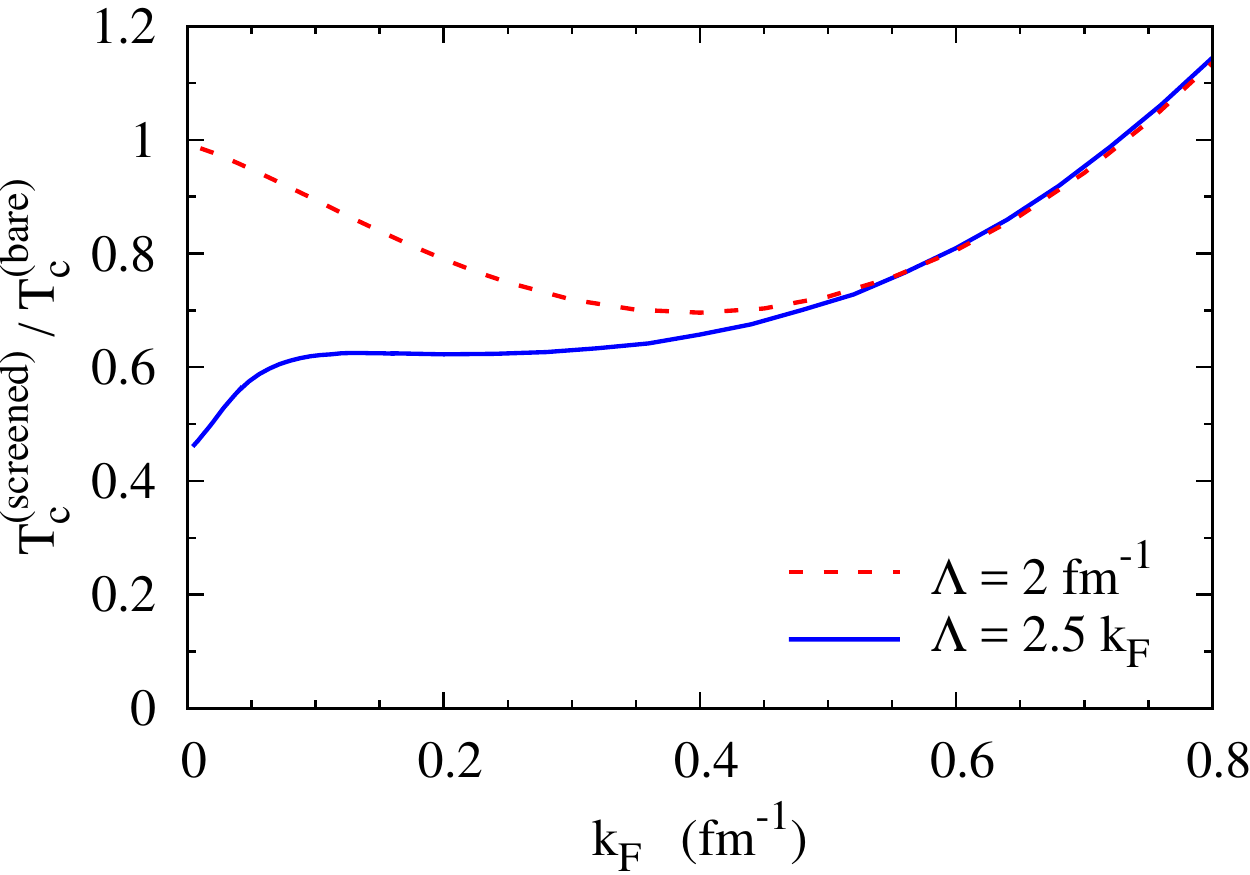}
\caption{\label{fig:Tcratio} Reduction of the critical temperature due
  to the screening correction $\Vind=V_a+V_b$ as a function of $k_F$,
  obtained with the constant cutoff $\Lambda=2\fmi$ (dashes) and with
  the density-dependent cutoff $\Lambda = 2.5\, k_F$ (solid line),
  respectively.}
\end{figure}

Since the results for $T_c$ obtained without the screening correction
is the same as the one we obtained before for $\Lambda = 2\fmi$, we
can concentrate on the correction of $T_c$ due to screening. In
\Fig{fig:Tcratio},
%%%%%%%%%%%%%%%%%%%%%%%%%%%%%%%%%%%%%%%%%%%%%%%%%%%%%%%%%%%%%%%%%%%%%%%%%%%%%%%%
%%%%%%%%%%%%%%%%%%%%%%%%%%%%%%%%%%%%%%%%%%%%%%%%%%%%%%%%%%%%%%%%%%%%%%%%%%%%%%%%
we therefore display the ratio of $T_c$ with screening to $T_c$
without screening as a function of $k_F$. The red dashes correspond to
the results shown already in \Fig{fig:Tc_vlowk}, obtained with a
constant cutoff $\Lambda = 2\fmi$, and we clearly see that the effect
of the screening correction vanishes at low density, as explained in
\Sec{sec:weak-coupling-fails}. The new results obtained with the
variable cutoff $2.5\, k_F$ are shown as the blue solid line. We see
that now the reduction of $T_c$ due to screening survives at low
densities, and in the limit $k_F\to 0$ it indeed seems to approach the
factor $(4e)^{-1/3} \approx 0.45$ predicted by GMB. Note that the
original GMB paper \cite{Gorkov1961} considers $k_F|a| \ll 1$, i.e.,
in the case of neutron matter, $k_F \ll 0.05 \fmi$.

%%%%%%%%%%%%%%%%%%%%%%%%%%%%%%%%%%%%%%%%%%%%%%%%%%%%%%%%%%%%%%%%%%%%%%%%%%%%%%%%
\section{Effect of the Nozi\`eres-Schmitt-Rink correction}
\label{sec:NSR}
%%%%%%%%%%%%%%%%%%%%%%%%%%%%%%%%%%%%%%%%%%%%%%%%%%%%%%%%%%%%%%%%%%%%%%%%%%%%%%%%
%%%%%%%%%%%%%%%%%%%%%%%%%%%%%%%%%%%%%%%%%%%%%%%%%%%%%%%%%%%%%%%%%%%%%%%%%%%%%%%%
\subsection{Brief summary of the formalism}
\label{sec:nsrformalism}
%%%%%%%%%%%%%%%%%%%%%%%%%%%%%%%%%%%%%%%%%%%%%%%%%%%%%%%%%%%%%%%%%%%%%%%%%%%%%%%%
In our previous work~\cite{Ramanan2013}, we had studied neutron matter
within the NSR approach using only the free-space renormalized
effective interaction $V_0$. In the present work, we will revisit the
inclusion of preformed pairs above $T_c$, including the induced
interaction $\Vind$ shown in \Fig{fig:diagrams}. For the sake of
completeness, we summarize briefly the key ideas and formulas of the
NSR approach. For more details, we refer the reader to
\Ref{Ramanan2013}.

Within the NSR approach, for a given chemical potential $\mu$, the
density of the interacting neutrons is enhanced by the pair
correlations that build up as a precursor effect to the superfluid
phase transition already above $T_c$. Therefore, the total density of
neutrons, $\rho_{\tot}$, can be written as
\begin{equation}
 \rho_{\tot} = \rho_0 + \rho_{\corr}\,.
 \label{eq:rhotot}
\end{equation}
The uncorrelated neutron density $\rho_0$ is given by
\begin{equation}
  \rho_0 = 2 \int\! \frac{d^3k}{(2\pi)^3}\, f(\xi(\vek{k}))\,,
\end{equation}	
where $f(\xi) = 1/(e^{\beta\xi}+1)$ is the Fermi-Dirac distribution
function (with $\beta = 1/T$) and the factor of $2$ arises due to the
spin degeneracy. The correlated density, $\rho_{\corr}$, in the
imaginary-time formalism~\cite{FetterWalecka}, is calculated to first
order in the single-particle self-energy $\Sigma$ as
\begin{multline}
  \rho_{\corr} = 2 \int\! \frac{d^3k}{(2\pi)^3} \frac{1}{\beta}
  \sum_{\omega_n} \big(\Gtemp_0(\vek{k}, \omega_n)\big)^2\\
  \times [\Sigma(\vek{k}, i\omega_n) - \re \Sigma(\vek{k},\xi(\vek{k}))]\,,
 \label{eq:rhocorr}
\end{multline}
where $\omega_n$ are the fermionic Matsubara frequencies and
$\Gtemp_0 = 1/(i\omega_n-\xi(\vek{k}))$ is the uncorrelated
single-particle Green's function. The subtraction of the on-shell
self-energy in the square bracket of \Eq{eq:rhocorr} is absent in the
original NSR approach. It takes into account the fact that
$\Gtemp_0$ includes already the in-medium quasiparticle energy
$\xi(\vek{k})$ which therefore must not be shifted by the self-energy
\cite{Zimmermann1985,Jin2010}.

Let us consider the first term without the
subtraction. $\Sigma(\vek{k}, i\omega_n)$ is calculated within the
ladder approximation, i.e.,
\begin{multline}
  \Sigma(\vek{k}, i\omega_n) = \int \frac{d^3 K}{(2 \pi)^2}
  \frac{1}{\beta} \sum_{\omega_N} \Gtemp_0(\vek{K} - \vek{k},
  \omega_N - \omega_n)\\
  \times \big\langle \tfrac{\vek{K}}{2} - \vek{k}\big|
    \Tmatrix(\vek{K}, i\omega_N)\big|\tfrac{\vek{K}}{2} - \vek{k}\big\rangle\,,
 \label{eq:sigma}
\end{multline}
where $\Tmatrix(\vek{K},i\omega_N)$ is the in-medium $\Tmatrix$ matrix
for the bosonic Matsubara frequency $\omega_N$ and total momentum
$\vek{K}$. The $\Tmatrix$-matrix is subsequently expanded in a partial
wave basis and we pick out only the $s$-wave contribution. Following
the steps outlined in~\cite{Ramanan2013} and analytically continuing to real
    $\omega$, one obtains for the correlated density within the NSR
    approach:
\begin{equation}
  \rho_{\corr,1} = -\frac{\partial}{\partial \mu} \int\!
  \frac{K^2 dK}{2\pi^2}\!
  \int\! \frac{d\omega}{\pi}\, g(\omega)
  \im\tr\log\big(1 - V \overline{G}_0^{(2)}\big)\,.
 \label{eq:rhocorr1}
\end{equation}
Here, $g(\omega) = 1/(e^{\beta \omega} - 1)$ is the Bose function, the
trace is taken w.r.t. the relative momentum $q$, $\overline{G}_0^{(2)}
= \overline{Q}(K,q)/(\omega-K^2/4m^*-q^2/m^*+2\mu)$ is the angle-averaged
(since we consider only the $s$ wave) retarded two-particle Green's
function, with $\overline{Q}(K,q)$ the Pauli-blocking factor
$1-f(\xi(\vek{K}/2-\vek{q}))-f(\xi(\vek{K}/2+\vek{q}))$ averaged over the
angle between $\vek{K}$ and $\vek{q}$. Working in the basis where
$V\overline{G}_0^{(2)}$ is diagonal, one can write \Eq{eq:rhocorr1} as
\begin{multline}
  \rho_{\corr,1} = -\frac{\partial}{\partial \mu} \int\!
  \frac{K^2 dK}{2\pi^2}\! \int\! \frac{d\omega}{\pi}\, g(\omega)\\
  \times \sum_{\nu} \im \log(1 - \eta_\nu(K, \omega))\,,
 \label{eq:rhocorr2}
\end{multline}
where $\eta_\nu$ are the (complex) eigenvalues of
$V\overline{G}_0^{(2)}$.

However, as mentioned below \Eq{eq:rhocorr}, one needs to correct for
the shift of the quasiparticle energies that comes from the real part
of the single-particle self-energy. Following~\cite{Ramanan2013},
we approximate $\Sigma(\vek{k},\xi(\vek{k}))$ by the first-order
(Hartree-Fock) self-energy and finally arrive at the following
correction: 
\begin{multline}
  \rho_{\corr,2} = \frac{\partial}{\partial \mu}
  \int\! \frac{K^2 dK}{2\pi^2}\, \frac{2}{\pi}\!\int\! q^2 dq\,
  g\big(\tfrac{K^2}{4m^*} + \tfrac{q^2}{m^*} - 2\mu\big)\\
    \times V(q,q)\overline{Q}(K,q),
	\label{eq:rhocorr_sub}
\end{multline}
which is added to \Eq{eq:rhocorr2}.

In \Ref{Ramanan2013}, the interaction $V$ that was used in
\Eqs{eq:rhocorr2} and~(\ref{eq:rhocorr_sub}) was the $\vlowk$
interaction obtained from AV$_{18}$ via the free-space renormalization
group evolution. But it seems straight-forward to include in addition
the medium corrections from diagrams (a) and (b), i.e., to use $V =
V_0+\Vind$. The only complication is that so far we calculated $\Vind$
only for a pair at rest, while we should now take into account the
finite center of mass momentum $\vek{K}$ of the pair.
%%%%%%%%%%%%%%%%%%%%%%%%%%%%%%%%%%%%%%%%%%%%%%%%%%%%%%%%%%%%%%%%%%%%%%%%%%%%%%%
\begin{figure}
\includegraphics[scale = 0.6]{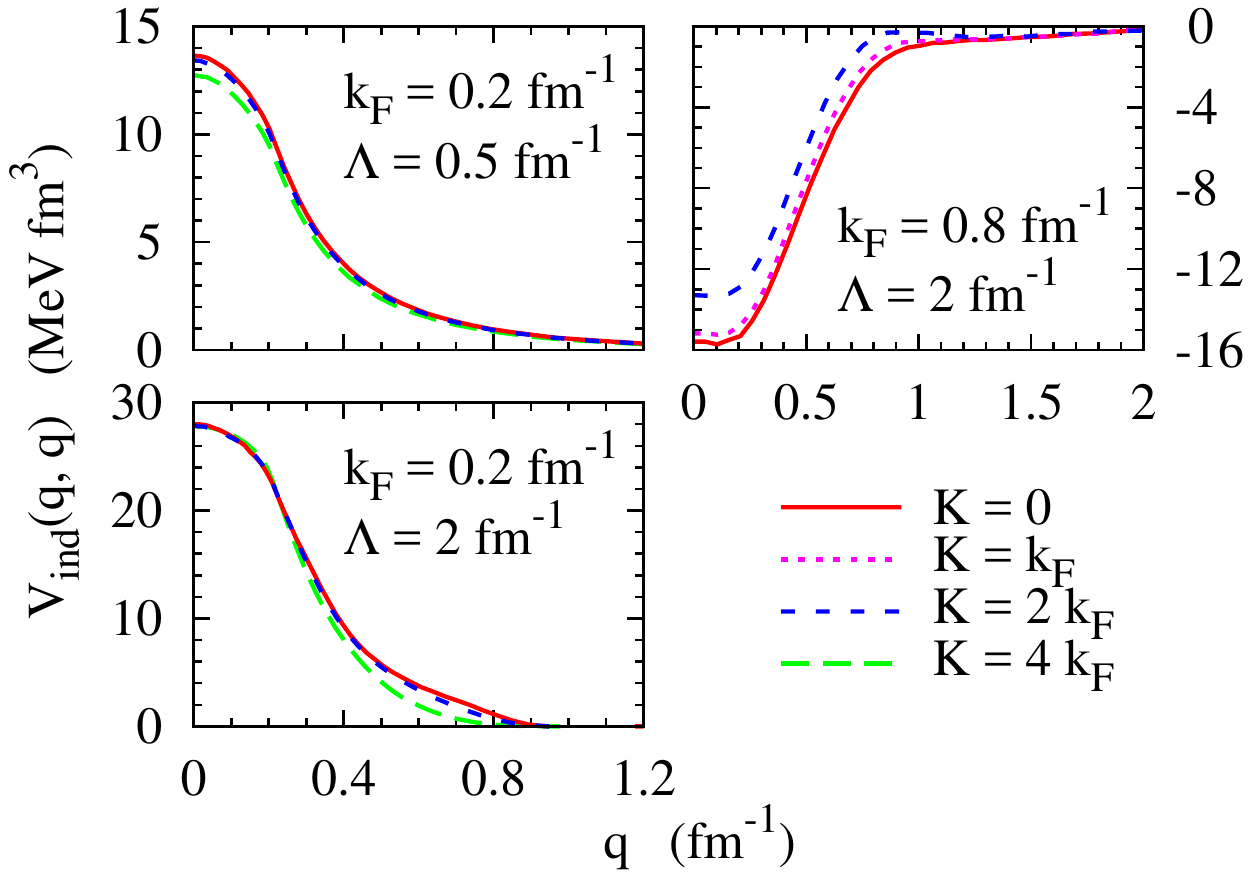}
\caption{Dependence of $\Vind$ on the momentum $K$ of the center of
  mass. For a density $k_F = 0.2 \, \fmi$ (left
    panels), the $K$-dependence is extremely weak even for momenta $K$
  exceeding $2k_F$. For $k_F = 0.8\fmi$ (upper right panel), the $K$
dependence is somewhat stronger but still too weak to make a
  significant contribution.}
\label{fig:Kdep}
\end{figure}
%%%%%%%%%%%%%%%%%%%%%%%%%%%%%%%%%%%%%%%%%%%%%%%%%%%%%%%%%%%%%%%%%%%%%%%%%%%%%%
	
To obtain the screening correction $\Vind$ for finite $\vek{K}$, some
minor modifications of \Eqs{eq:diaga} and (\ref{eq:diagb}) are
necessary. Details are given in Appendix \ref{app:Kdep}. We have
checked that, at least for $T=T_c$, the contributions to the integrals
in \Eqs{eq:rhocorr2} and (\ref{eq:rhocorr_sub}) come only from
$K\lesssim 2k_F$. As seen in \Fig{fig:Kdep}, numerically it turns out 
that the $K$ dependence of
$\Vind$ is very weak for $K<2k_F$ in the range of $k_F$ where the NSR
correction can be expected to be important. We will therefore neglect
this $K$ dependence and use in \Eqs{eq:rhocorr2} and
(\ref{eq:rhocorr_sub}), the screening correction calculated for $K=0$.

There are a couple more points that need to be discussed. For
instance, now one has two different densities, the uncorrelated one,
$\rho_0$, and the corrected one, $\rho_{\tot}$. The question arises
which density one should use in the calculation of the induced
interaction $\Vind$. Since $\Vind$ is computed with uncorrelated
propagators and occupation numbers, it seems more appropriate to take
only the uncorrelated density $\rho_0$ into account in the calculation
of $\Vind$. From the derivation of \Eqs{eq:rhocorr2} and
(\ref{eq:rhocorr_sub}) it is also clear that the derivatives
$\partial/\partial\mu$ should be taken with the interaction $\Vind$
kept constant (and the effective mass $m^*$, too). This points to
fundamental problems of the present approach, which is clearly not a
fully consistent treatment of both particle-particle and particle-hole
fluctuations. Nevertheless, we expect to get at least a rough idea
about the change of the NSR effect when the pair
correlations are modified by screening.
%%%%%%%%%%%%%%%%%%%%%%%%%%%%%%%%%%%%%%%%%%%%%%%%%%%%%%%%%%%%%%%%%%%%%%%%%%%%%%%%
\subsection{Results}
\label{sec:res_nsr}
\begin{figure}
  \includegraphics[scale=0.31]{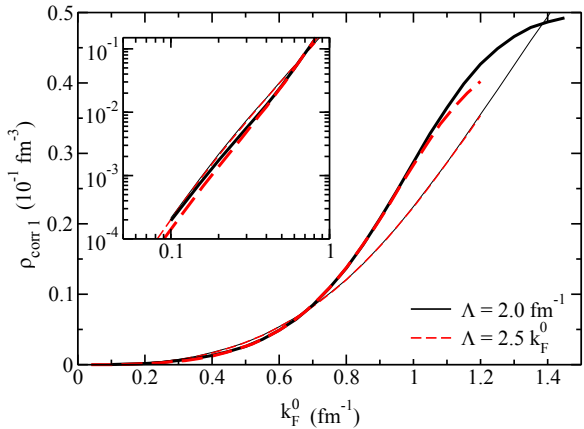}
  \caption{The un-subtracted correlated density $\rho_{\corr,\,1}$ as
    a function of the Fermi momentum $k_F^0$ with and without the
    screening correction, calculated at the respective critical
    temperatures $T_c$. Here, the black solid lines and the red dashed
    lines show the results for the two cutoffs of $\Lambda = 2.0\,
    \fmi$ and $\Lambda = 2.5 \, k_F^0$. The thin lines contain only
    $V_0$, while the thick lines include the induced interactions.
    The inset in the figure magnifies the cutoff dependence in
    $\rho_{\corr, \,1}$ at low densities. The Fermi-liquid parameters
    are calculated using the SLy4 interaction.
    \label{fig:rhocorr1}}
\end{figure}

%%%%%%%%%%%%%%%%%%%%%%%%%%%%%%%%%%%%%%%%%%%%%%%%%%%%%%%%%%%%%%%%%%%%%%%%%%%%%%%%
Before discussing the critical temperature as a function of density,
let us look at the density correction. The un-subtracted correlated
density, $\rho_{\corr,1}$ as a function of the Fermi-momentum
corresponding to the uncorrelated density $\rho_0$, denoted here as
$k_F^0 = (3 \pi^2 \rho_0)^{1/3}$, is shown in \Fig{fig:rhocorr1}.
%%%%%%%%%%%%%%%%%%%%%%%%%%%%%%%%%%%%%%%%%%%%%%%%%%%%%%%%%%%%%%%%%%%%%%%%%%%%%%%%
%%%%%%%%%%%%%%%%%%%%%%%%%%%%%%%%%%%%%%%%%%%%%%%%%%%%%%%%%%%%%%%%%%%%%%%%%%%%%%%%
The black solid lines and the red dashed lines represent two different
cutoff choices, a constant cutoff $\Lambda = 2.0 \, \fmi$ and a
density dependent cutoff $\Lambda = 2.5\, k_F^0$. The thin lines show
the correlated density $\rho_{\corr,1}$ with only the free-space
interaction $V_0$. Analogous to Fig. 5 of \Ref{Ramanan2013}, we see
that $\rho_{\corr,1}$ with only $V_0$ is independent of the
cutoff. With the inclusion of the induced interaction (thick lines) we
note that the cutoff dependence of $\rho_{\corr,1}$ is again
negligible, except at very low densities (see inset), where we found
stronger screening with the variable cutoff compared to the fixed
cutoff (cf. \Fig{fig:Tcratio}). In addition, up to $k_F^0 \sim 0.7 \,
\fmi$, the correlated density $\rho_{\corr,1}$ with the induced
interaction is smaller than the correlated density without the induced
interaction, consistent with the earlier observation that the induced
interaction screens $V_0$. However, in the range of Fermi-momenta
where the induced interaction anti-screens $V_0$, the correlated
density $\rho_{\corr,1}$ is larger than the corresponding quantity
without the induced interaction.

\begin{figure}
\includegraphics[scale=0.31]{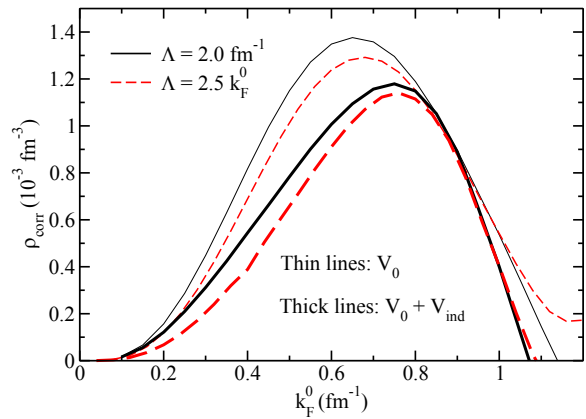}
\caption{Subtracted correlated density $\rho_\corr$ as a function of
  $k_F^0$ with and without screening, calculated at the respective
  critical temperatures. See \Fig{fig:rhocorr1} for details.
  \label{fig:rhocorr}}
\end{figure}

Let us now turn our attention to the correlated density with the
first-order (Hartree-Fock) subtraction, $\rho_{\corr}$. The dependence
of $\rho_{\corr}$ on $k_F^0$ is shown in \Fig{fig:rhocorr}.
%%%%%%%%%%%%%%%%%%%%%%%%%%%%%%%%%%%%%%%%%%%%%%%%%%%%%%%%%%%%%%%%%%%%%%%%%%%%%%%%
%%%%%%%%%%%%%%%%%%%%%%%%%%%%%%%%%%%%%%%%%%%%%%%%%%%%%%%%%%%%%%%%%%%%%%%%%%%%%%%%
As in \Fig{fig:rhocorr1}, the black solid lines and the red dashed
lines show results for the two different cutoffs: the constant cutoff
$\Lambda = 2.0 \, \fmi$ and the density dependent cutoff $\Lambda =
2.5\, k_F^0$, respectively. For low $k_F^0$, we see that the
correlated density with the inclusion of the induced interaction
(thick lines) is smaller than in the $V_0$-only case (thin lines)
which is consistent with the screening of $V_0$ by $\Vind$ and similar
to the trend seen in \Fig{fig:rhocorr1}. However, what is surprising
is that even in the region where $\Vind$ anti-screens $V_0$, the
correlated density gets smaller with the inclusion of $\Vind$ compared
to the $V_0$-only case.  Further, one notices strong cutoff dependence
in the low $k_F^0$ region if one compares the solid black line with
the red dashed line, both with and without the inclusion of the
induced interaction. Both these observations are completely different
from \Fig{fig:rhocorr1} and are clearly the effect of the Hartree-Fock
subtraction. For the density dependent cutoff, at low-densities, this
subtraction should get better as the interaction gets more
perturbative at smaller cutoffs. However, at high densities, where the
subtraction $\rho_{\corr,2}$ is almost of the same magnitude as
$\rho_{\corr,1}$ itself, the Hartree-Fock approximation is not precise
enough to give a reliable result for the subtracted
$\rho_{\corr}$. Hence, the suppression of the correlated density for
higher $k_F^0$ in \Fig{fig:rhocorr}, once the induced interaction is
included, is probably unphysical. Fortunately, in this region,
$\rho_{\corr}$ is completely negligible compared to $\rho_0$.

\begin{figure}
  \includegraphics[scale = 0.31]{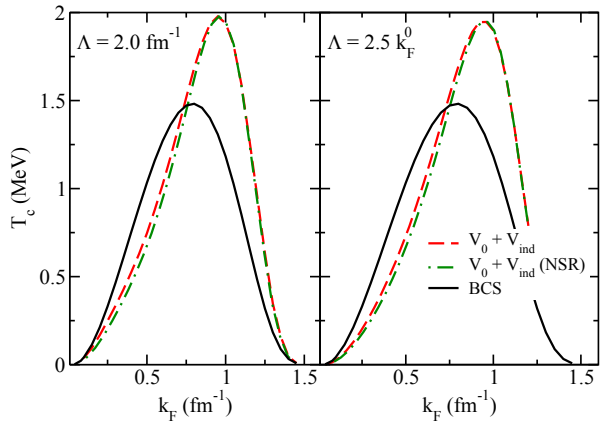}
  \caption{$T_c$ versus $k_F$: (Left panel) Results with fixed cutoff
    $\Lambda = 2\fmi$; (Right panel) density dependent cutoff $2.5\,k_F$. The
    green dashed-dotted lines are the full results including the
    induced interaction $\Vind$ and the correlated density
    $\rho_{\corr}$ in the NSR framework. For comparison, we also show
    the BCS result (only $V_0$ and $\rho_0$, black solid lines) and
    the results obtained with the induced interaction $\Vind$ but
    without the NSR correction (red dashed lines).
	\label{fig:Tcvskf}}
\end{figure}

Now we are in the position to discuss the final results for the
critical temperature $T_c$ as a function of $k_F$, displayed in
\Fig{fig:Tcvskf}.
%%%%%%%%%%%%%%%%%%%%%%%%%%%%%%%%%%%%%%%%%%%%%%%%%%%%%%%%%%%%%%%%%%%%%%%%%%%%%%%%
%%%%%%%%%%%%%%%%%%%%%%%%%%%%%%%%%%%%%%%%%%%%%%%%%%%%%%%%%%%%%%%%%%%%%%%%%%%%%%%%
Note that in the NSR framework, $T_c$ as a function of $\mu$ is
computed as usual, and only the relation between $\mu$ and $k_F$ (and
$\rho$) is changed. Here, $k_F$ denotes the Fermi momentum
corresponding to the total density including $\rho_{\corr}$, i.e.,
$k_F = (3\pi^2\rho_{\tot})^{1/3}$ (green dashed-dotted lines). As a
consequence, the presence of the correlated density $\rho_{\corr}$
shifts the curve slightly to the right. In order to make easy
comparisons, we also show the BCS result (solid line) and the results
obtained with $\Vind$ but without the NSR correction (red dashed
lines). In both panels, we note that the pair correlations lower the
transition temperature compared to the one with screening alone at the
same $k_F$. However, the trends already observed with the medium
corrections (\Figs{fig:Tc_vlowk} and \ref{fig:Tcratio}), i.e.,
screening at low densities and anti-screening at high densities,
remain unchanged, since the NSR effect is much weaker than the
screening or anti-screening effect of $\Vind$.

Please notice that the relation $\Delta_{T=0}(k_F) = 1.76\, T_c$ for a
  given $k_F$, mentioned in \Sec{sec:tc}, is \textit{not} valid for
  the NSR results.
%%%%%%%%%%%%%%%%%%%%%%%%%%%%%%%%%%%%%%%%%%%%%%%%%%%%%%%%%%%%%%%%%%%%%%%%%%%%%%%%
\section{Conclusions}
\label{sec:conclude}
%%%%%%%%%%%%%%%%%%%%%%%%%%%%%%%%%%%%%%%%%%%%%%%%%%%%%%%%%%%%%%%%%%%%%%%%%%%%%%%%
It has been known for a long time that screening corrections have a
very strong effect on the superfluid transition temperature of neutron
matter. Also the fact that the RPA, diagram (b), reduces the effect of
diagram (a), has been known before \cite{Cao2006}. However, in
\Ref{Cao2006} the effect of diagram (b) was too weak to overcome the
strong screening generated by diagram (a), while we find that, around
$n\gtrsim 0.01-0.02\fmithree$, the net effect of $\Vind$ is attractive
and screening turns into anti-screening. A similar effect was found in
\Ref{Schulze1996}, but only at much higher densities ($n\gtrsim
0.07\fmithree$). There are three main differences between our
calculation and that of \Ref{Cao2006}. First, we are using $\vlowk$
while in \cite{Cao2006} the Br\"uckner $G$ matrix was used in the
vertices. Second, while we keep the full momentum dependence of the
non-local interaction, the vertices in \cite{Cao2006} were replaced by
an average matrix element. Probably the most important difference,
however, is the choice of the Landau parameters. Here, we take them
from a phenomenological energy density functional (SLy4). Since this
functional was fitted to  QMC results
for the neutron matter equation of state, we assume that the Landau
parameters are rather well determined. The anti-screening effect
arises primarily from the enhancement of the attractive density
($S=0$) fluctuations due to the strongly negative $f_0$ parameter. In
\cite{Cao2006}, on the contrary, the Landau parameters were obtained
following the so-called Babu-Brown theory as explained in
\cite{Schulze1996}. This results in particular in a much smaller (less
negative) value of the $f_0$ parameter, and as a consequence, the
density fluctuations are not strong enough to compensate for the repulsive
effect of the spin-density ($S=1$) fluctuations.

We addressed in some detail the problem of the low density limit. When
a constant (density-independent) potential $V$ is used in the vertices
of diagram (a), the screening effect disappears at low density,
although from the weak-coupling formula one would conclude that the
gap should be reduced by the factor $(4e)^{-1/3}$ predicted by GMB
\cite{Gorkov1961}. We explained why the weak-coupling formula
fails in this particular case. We then observed that GMB used the full
$\calT$ matrix instead of the potential $V$ in the vertices of diagram
(a). This allowed us to finally recover the GMB result, namely by
using for each density a $\vlowk$ interaction evolved to a cutoff
$\Lambda$ that scales with $k_F$. In this way, one ensures that, on
the one hand, one does not cut the relevant degrees of freedom
($q\lesssim k_F$), and on the other hand, the Born term is already a
reasonable approximation to the full $\calT$ matrix at momenta of the
order of $q\sim k_F$.

In the last part of the paper we discussed the effect of preformed
pairs on the critical temperature $T_c$ in the NSR framework. In spite
of some cutoff and regulator dependence in the detailed study of the
correlated density $\rho_{\corr}$, one can clearly see that due to
$\rho_{\corr}$ the critical temperature $T_c$ for a given density is
slightly reduced. But this effect is much less important than the
induced interaction. Compared to ultracold atoms in the unitary limit
or even on the BEC ($a>0$) side of the BCS-BEC crossover, neutron
matter remains more or less in a weakly coupled regime at all
densities.

There remain obviously many open questions. For instance, as discussed
in \cite{Cao2006}, the reduction of the quasiparticle residue $Z<0$
can lead to a reduction of $T_c$, and this
effect has not been included in the present study. Another point that
clearly needs to be improved is the Landau approximation in the
RPA. In principle, it is only valid for momentum transfer $k\ll k_F$,
but in the induced interaction, the relevant range of momentum
transfers is $0\leq k\leq 2k_F$. In the framework of Skyrme
interactions it is actually straight-forward to solve the RPA beyond
the Landau approximation,
and this issue will be addressed in a future study.

Concerning the meaning of the density dependent cutoff introduced in
\Sec{sec:ddcutoff}, one might wonder how this is related to the
so-called functional renormalization-group approach in which one
solves flow equations in the medium, integrating out all momenta
except the Fermi surface. Such approaches have been used to include
screening corrections in a non-perturbative way for neutron matter
\cite{Schwenk2003} and ultracold atoms
\cite{Gubbels2008,Floerchinger2008}. In the context of the small
cutoff, one should also mention that lowering the cutoff induces
three- and higher-body interactions. These are neglected in $\vlowk$
since it is obtained for two particles in free space. A better
approach in this respect would be the in-medium similarity
renormalization group \cite{Hergert2017}, which allows one to include
many-body effects at least approximately into the effective two-body
interaction.

Because of the extreme sensitivity of the gap and the critical
temperature to the details of the effective interaction, it seems
likely that large theoretical uncertainties will remain. Maybe
astrophysical observations of neutron stars can help to decide which
theory is correct.

%%%%%%%%%%%%%%%%%%%%%%%%%%%%%%%%%%%%%%%%%%%%%%%%%%%%%%%%%%%%%%%%%%%%%%%%%%%%%%%%
\appendix
%%%%%%%%%%%%%%%%%%%%%%%%%%%%%%%%%%%%%%%%%%%%%%%%%%%%%%%%%%%%%%%%%%%%%%%%%%%%%%%%
\section{Partial wave expansion of the Gogny force}
\label{app:Gogny}
%%%%%%%%%%%%%%%%%%%%%%%%%%%%%%%%%%%%%%%%%%%%%%%%%%%%%%%%%%%%%%%%%%%%%%%%%%%%%%%%
We expand the Gogny force as given in Ref.~\cite{Decharge1980} into
partial waves, neglecting the spin-orbit term as in
\cite{Shen2005}. The resulting matrix elements in the $nn$
channel read:
\begin{multline}
  \langle Q|V_{ls}|Q'\rangle = \frac{1}{4\pi}\sum_{i=1,2}[W_i-H_i+(-1)^S(M_i-B_i)]\\
  \times (\sqrt{\pi}\mu_i)^3
  e^{-(Q^2+Q^{\prime\, 2})\mu_i^2/4} i_l(QQ'\mu_i^2/2) 
\end{multline}
where $i_l(z) = \sqrt{\pi/2z} I_{l+1/2}(z)$ is a modified spherical
Bessel function of the first kind \cite{AbramowitzStegun}: $i_0(z) =
\sinh(z)/z$, etc. The antisymmetrized matrix elements are then
obtained by $\langle Q|\tilde{V}_{ls}|Q'\rangle = [1+(-1)^{l+s}]
\langle Q|V_{ls}|Q'\rangle$. The density dependent contact term of the Gogny
force does not contribute since it acts only in the neutron-proton
channel.

Concerning the values of $\mu_i$, $W_i$, $H_i$, $B_i$, and $M_i$, we
use either the parameterization D1 \cite{Decharge1980} to compare with
Ref.~\cite{Shen2005} or the more recent parameterization D1N
\cite{Chappert2008}.

%%%%%%%%%%%%%%%%%%%%%%%%%%%%%%%%%%%%%%%%%%%%%%%%%%%%%%%%%%%%%%%%%%%%%%%%%%%%%%%%
\section{Fermi-liquid parameters}
\label{app:Landau}
%%%%%%%%%%%%%%%%%%%%%%%%%%%%%%%%%%%%%%%%%%%%%%%%%%%%%%%%%%%%%%%%%%%%%%%%%%%%%%%%
In this work, we use the Fermi-liquid parameters from the SLy4
parameterization of the Skyrme functional \cite{Chabanat1997} or from
the D1N parametrization of the Gogny force \cite{Chappert2008}. The
explicit expressions in terms of the Skyrme-force parameters $t_i$,
$x_i$ ($i=0\dots 3$), and $\sigma$ read \cite{Margueron2002}
\begin{align}
  \frac{1}{m^*} =& \frac{1}{m}+
    \tfrac{1}{4}[t_1(1-x_1)+3t_2(1+x_2)]\rho\,,\\
  f_0 =& \tfrac{1}{2}t_0(1-x_0)
    +\tfrac{1}{4}[t_1(1-x_1)+3t_2(1+x_2)]k_F^2\nonumber\\
    &+\tfrac{1}{24}t_3(1-x_3)(1+\sigma)(2+\sigma)\rho^\sigma\,,\\
  g_0 = &\tfrac{1}{2}t_0(x_0-1)
    +\tfrac{1}{4}[t_1(x_1-1)+t_2(1+x_2)]k_F^2\nonumber\\
    &+\tfrac{1}{12}t_3(x_3-1)\rho^\sigma\,.
\end{align}
In the case of the Gogny force, one obtains the following expressions
for the Fermi-liquid parameters \cite{Shen2005}:
\begin{align}
  \frac{1}{m^*} =& \frac{1}{m}+\frac{m}{\sqrt{\pi}k_F}\sum_{i=1,2}
  \mu_i(W_i+2B_i-H_i-2M_i)\nonumber\\
  &\times z_i e^{-z_i} i_1(z_i)\,,\\
  f_0 =& \sum_{i=1,2} \frac{(\sqrt{\pi}\mu_i)^3}{2}[(2W_i+B_i-2H_i-M_i)\nonumber\\
      &-(W_i+2B_i-H_i-2M_i)e^{-z_i} i_0(z_i)]\,,\\
  g_0 =& \sum_{i=1,2} \frac{(\sqrt{\pi}\mu_i)^3}{2}[(B_i-M_i)
      -(W_i-H_i)e^{-z_i} i_0(z_i)]\,,
\end{align}
where $z_i = k_F^2\mu_i^2/2$.
%%%%%%%%%%%%%%%%%%%%%%%%%%%%%%%%%%%%%%%%%%%%%%%%%%%%%%%%%%%%%%%%%%%%%%%%%%%%%%%%
\section{Angle-averaged Lindhard function}
\label{app:Piav}
%%%%%%%%%%%%%%%%%%%%%%%%%%%%%%%%%%%%%%%%%%%%%%%%%%%%%%%%%%%%%%%%%%%%%%%%%%%%%%%%
For $q,q'\neq 0$, the general explicit expression for the
angle-averaged Lindhard function defined in \Eq{eq:Piav} reads
\begin{multline}
\langle\tilde{\Pi}_0\rangle = -\frac{1}{3}
+\frac{k_F^2}{48 qq'}[F(2-x_-)+F(2+x_-)\\
  -F(2-x_+)-F(2+x_+)]\,,
\end{multline}
with $F(x) = x^2(6-x)\ln|x|$ and $x_{\pm} = |q\pm q'|/k_F$. In the
special case of interest $q=q'=k_F$ mentioned in the main text this
gives $\langle\tilde{\Pi}_0\rangle = -\tfrac{1}{3}\ln 4e
\approx -0.795$. The expression for the cases $q\neq q'=0$ or
$q'\neq q=0$ reads
\begin{equation}
  \langle\tilde{\Pi}_0\rangle
    = \frac{x^2-4}{8x} \artanh\Big(\frac{x}{2}\Big)-\frac{1}{2}\,,
\end{equation}
with $x = q/k_F$ or $q'/k_F$, respectively. In the special case
$q=q'=0$, one obtains $\langle\tilde{\Pi}_0\rangle = -1$. For $q\gg
k_F$ or $q'\gg k_F$, $\langle\tilde{\Pi}_0\rangle$ tends to
zero.
%%%%%%%%%%%%%%%%%%%%%%%%%%%%%%%%%%%%%%%%%%%%%%%%%%%%%%%%%%%%%%%%%%%%%%%%%%%%%%%%
\section{Computation of the screening corrections for pairs with finite
  total momentum}
\label{app:Kdep}
%%%%%%%%%%%%%%%%%%%%%%%%%%%%%%%%%%%%%%%%%%%%%%%%%%%%%%%%%%%%%%%%%%%%%%%%%%%%%%%%
In \Fig{fig:diagrams} and the corresponding \Eqs{eq:diaga} and
(\ref{eq:diagb}), we have considered from the beginning a pair at
rest (with respect to the medium). However, for the NSR correction,
one needs pairs with finite total momentum $\vek{K}$. In order to
compute the screening corrections $\Vind=V_a+V_b$ for $\vek{K}\neq 0$, one
has to change the definitions of the vectors $\vek{Q}_1$,
$\vek{Q}_1^\prime$, $\vek{Q}_2$, and $\vek{Q}_2^\prime$ that appear in
\Eqs{eq:diaga} and (\ref{eq:diagb}). For diagram (a),
one has to replace \Eq{definition-Q-a} by
\begin{equation}
\begin{split}
  &\vek{Q}_1 = \frac{\vek{q}+\vek{p}}{2}-\frac{\vek{K}}{4}\,,\quad
  \vek{Q}'_1 = \frac{\vek{q}'-\vek{k}+\vek{p}}{2}-\frac{\vek{K}}{4}\,,\\
  &\vek{Q}_2 = \frac{\vek{q}+\vek{k}-\vek{p}}{2}+\frac{\vek{K}}{4}\,,\quad
  \vek{Q}'_2 = \frac{\vek{q}'-\vek{p}}{2}+\frac{\vek{K}}{4}\,.
  \label{definition-Q-a-finiteK}
\end{split}
\end{equation}
For diagram (b), the definition (\ref{definition-Q-b}) has to be replaced by
\begin{equation}
\begin{split}
  &\vek{Q}_1 = \frac{\vek{q}+\vek{p}_1}{2}-\frac{\vek{K}}{4}\,,\quad
  \vek{Q}'_1 = \frac{\vek{q}'-\vek{k}+\vek{p}_1}{2}-\frac{\vek{K}}{4}\,, \\
  &\vek{Q}_2 = \frac{\vek{q}+\vek{k}-\vek{p}_2}{2}+\frac{\vek{K}}{4}\,,\quad
  \vek{Q}'_2 = \frac{\vek{q}'-\vek{p}_2}{2}+\frac{\vek{K}}{4}\,.
  \label{definition-Q-b-finiteK}
\end{split}
\end{equation}
However, for diagram (b), this is not sufficient, because we used the
isotropy to replace the sum over the three spin projections $m_S =
-1,0,1$ of the $S=1$ particle-hole excitation by the contribution of
$m_S=0$, multiplied by three. But for $\vek{K}\neq 0$, the isotropy is
lost and therefore the contributions of the three spin projections
will not be equal any more. Nevertheless, after summation over $m_S$,
the final result for $V_b$ can only depend on $K=|\vek{K}|$ and not on
the direction of $\vek{K}$. Hence, we can average over the angle of
$\vek{K}$. By doing so, we have restored the isotropy and it is
therefore again sufficient to compute only the contribution of $m_S=0$
and to multiply the result by three.
%%%%%%%%%%%%%%%%%%%%%%%%%%%%%%%%%%%%%%%%%%%%%%%%%%%%%%%%%%%%%%%%%%%%%%%%%%%%%%%%

\end{document}